\documentclass[aps,showpacs,superscriptaddress,nofootinbib,amsmath,amssymb,10pt]{revtex4}
 \usepackage{graphicx}
\usepackage{dcolumn}
\usepackage{bm}
\usepackage{amsmath}
\usepackage{amssymb}
\usepackage{enumerate}
\usepackage{xcolor}

\graphicspath{./fig}
\begin{document}
 \title{Nuclear medium effects in the deep inelastic $\nu_\tau/\bar\nu_\tau-{}^{40}$Ar scattering at DUNE energies}
\author{F. Zaidi}
\affiliation{Department of Physics, Aligarh Muslim University, Aligarh - 202002, India}
\author{V. Ansari}
\affiliation{Department of Physics, Aligarh Muslim University, Aligarh - 202002, India}
\author{M. Sajjad Athar\footnote{Corresponding author: sajathar@gmail.com}}
\affiliation{Department of Physics, Aligarh Muslim University, Aligarh - 202002, India}
\author{H. Haider}
\affiliation{Department of Physics, Aligarh Muslim University, Aligarh - 202002, India}
\author{I. \surname{Ruiz Simo}}
\affiliation{Departamento de F\'{\i}sica At\'omica, Molecular y Nuclear,
and Instituto de F\'{\i}sica Te\'orica y Computacional Carlos I,
Universidad de Granada, Granada 18071, Spain}
\author{S. K. Singh}
\affiliation{Department of Physics, Aligarh Muslim University, Aligarh - 202002, India}

\begin{abstract}
The nuclear medium effects are studied in the $\nu_\tau/\bar\nu_\tau$ interactions from nuclei in the 
 deep inelastic scattering (DIS) region and 
applied to the $^{40}Ar$ nucleus to obtain the scattering cross sections in the energy region of the proposed DUNE experiment.
The free nucleon structure functions ($F_{iN}(x,Q^2);~(i=1-5)$) have been calculated at the next-to-leading order (NLO) 
using Martin-Motylinski-Harland Lang-Thorne 2014 as well as the Coordinated Theoretical-Experimental Project on QCD parameterizations for parton distribution functions (PDFs) and including the effect of
perturbative and nonperturbative QCD corrections~\cite{Ansari:2020xne}. These free nucleon structure functions are then convoluted with the nucleon spectral function in the nucleus to
obtain the nuclear structure functions ($F_{iA}(x,Q^2);~(i=1-5)$). The nucleon spectral function takes into account the Fermi motion and the binding energy of the
nucleons as well as the nucleon correlations within the nucleus. These nuclear structure functions are then used to calculate the deep inelastic
scattering cross sections. Moreover, the contribution of $\pi$ and $\rho$ mesons as well as the corrections due to the shadowing and 
antishadowing effects in the relevant kinematic region of the Bjorken variable $x$ are also included.  
The numerical results for the nuclear structure functions and scattering cross sections have been presented and compared with the results obtained in the phenomenological
approach using nuclear PDFs from nCTEQ15 and nCTEQnu.
  \end{abstract}
\maketitle
\section{Introduction}
The tau neutrino ($\nu_\tau$) is experimentally the least studied Standard Model lepton due to the inherent difficulties in producing a $\nu_\tau$ beam in the laboratory.
The ``Direct Observation of the NU Tau'' (DONUT) collaboration was the first to directly observe 
the tau neutrino charge current interaction in their experiment~\cite{Nakamura:1999dp}.  Later attempts have been made by the ``Neutrino Oscillation MAgnetic Detector'' (NOMAD)~\cite{Astier:2001yj} 
and the ``Oscillation Project with Emulsion-tRacking Apparatus'' (OPERA)~\cite{Agafonova:2018auq,Agafonova:2015jxn,Agafonova:2014ptn,Agafonova:2014bcr} 
collaboration experiments to study the $\nu_\tau-$nucleon charged current interactions by producing $\tau-$lepton generated through the $\nu_\mu \rightarrow \nu_\tau$ oscillations in the $\nu_\mu$ beam available at the high
energy accelerators. The NOMAD collaboration~\cite{Astier:2001yj} 
has observed 9 $\nu_\tau$ events and the OPERA collaboration~\cite{Agafonova:2018auq} has observed 10 $\nu_\tau$ events. On the other hand, $\nu_\tau$ induced  $\tau-$lepton production has 
also been reported in the 
atmospheric neutrino sector by the SuperK~\cite{Abe:2012jj,Li:2017dbe} and IceCube~\cite{Aartsen:2019tjl} collaborations, where $338\pm72.7(stats\pm sys)$ and 934 tau leptons have been observed, 
respectively using the $\nu_\tau$ beam from the $\nu_\mu \rightarrow \nu_\tau$ oscillations in the energy region of 3.5 GeV $< E_{\nu_\tau} <$70 GeV and 5.6 GeV$< E_{\nu_\tau} 
<$56 GeV. The corresponding cross sections have been reported to be 
$(0.94\pm0.20)\times 10^{-38}$ cm$^2$ in the energy region of 3.5 GeV$<E_{\nu_\tau}<70$ GeV by the SuperK collaboration\cite{Li:2017dbe} and $\sigma^{const}_{\nu_\tau}=(0.39\pm0.13\pm0.13)\times10^{-38}$ cm$^2/$GeV for 
$E_{\nu_\tau}<300$ GeV by the DONUT collaboration~\cite{Kodama:2007aa}. Future experiments with the atmospheric neutrinos are also proposed to be performed by the HyperK 
collaboration~\cite{Hadley:2016jpp} with a larger volume of the ultra pure water target which is almost an order of magnitude larger than the SuperK detector-target~\cite{Abe:2012jj,Li:2017dbe}.

In the accelerator sector, some experiments are planning to use $\nu_\tau$ beams from the $\nu_\mu \rightarrow \nu_\tau$ oscillations as well as from the decay of
$D_s-$mesons ($D_s \rightarrow \tau \; \nu_\tau$) which are produced in the high energy proton-nucleus collisions. For example, the  Search for Hidden Particles (SHiP) 
collaboration at CERN~\cite{SHiP:2018xqw, DiCrescenzo:2016irr} and the Deep Underground Neutrino Experiment (DUNE) collaboration at Fermilab~\cite{Abi:2020qib,Abi:2018dnh,Abi:2020mwi}  plan 
to use the $\nu_\tau$ beam from the $\nu_\mu \rightarrow \nu_\tau$ oscillations, while the 
DsTau collaboration~\cite{Aoki:2019jry} plans to use $\nu_\tau$ beam from the decays of $D_s$ mesons. 
Recently at CERN, the FASER$\nu$ experiment has been proposed to detect collider neutrinos using emulsion detector~\cite{Jodlowski:2020vhr}.
All these experimental proposals planned and approved to be performed, as well as the earlier experiments, use nuclear targets to study the $\nu_\tau-$nucleon interactions. In Table~\ref{Table-1}, we give the list of various nuclear targets used or to be used in these experiments.
The extraction of $\nu_\tau-$nucleon interaction observables like the total and differential scattering cross sections as well as the oscillation parameters in the $\nu_\tau$ sector would have systematic uncertainty arising due to the model dependence
of the $\nu_\tau-$nucleus cross sections in treating the nuclear medium effects. This will be in addition to the uncertainties 
in high energy $\nu_\tau-$nucleon cross sections present inherently in 
the case of free nucleon targets, some of which are discussed in the literature~\cite{Paschos:2001np,Jeong:2010nt,Hagiwara:2003di,Conrad:2010mh,Ansari:2020xne}. 
\begin{table}[h]\label{Table-1}
 \centering
 \begin{tabular}{|c|c|c|c|c|c|c|c|c|c|}\hline\hline
  Experiment&NOMAD&DONuT&OPERA&DUNE&SHiP&DsTau&SuperK&HyperK&FASER$\nu$\\
  Nuclear Target&$^{56}Fe$&Emulsion nuclei&$^{208}Pb$&$^{40}Ar$&$^{208}Pb$&$^{208}Pb$&$^{16}O$&$^{16}O$&Emulsion nuclei\\\hline\hline
 \end{tabular}
 \caption{$\nu_\tau$ experiments with nuclear targets.}
\end{table}
 
The need for studying the nuclear medium effects in the $\nu_\tau-$nucleus interactions has been emphasized earlier in some experimental and theoretical papers in the context of $\tau-$lepton production 
induced by the tau neutrinos~\cite{Paschos:2001np,Conrad:2010mh}. 
With the increasing interest in $\nu_\tau$ physics~\cite{Aartsen:2019tjl, Abi:2020qib,Abi:2018dnh,Abi:2020mwi, Aoki:2019jry, SHiP:2018xqw, DiCrescenzo:2016irr, Jodlowski:2020vhr} it is also important to understand the $\nu_\tau-$ interaction cross sections in the nuclear targets~\cite{tau2021, nutau2021}.
But a serious attempt to study these effects quantitatively has been lacking except for the earlier work of Paschos and Yu~\cite{Paschos:2001np} in 
which the nuclear medium effects in the $\nu_\tau-$nucleus scattering in the deep inelastic region has been incorporated at the leading order in the massless limit of quarks using the 
phenomenological nuclear structure functions of Eskola et al.~\cite{Eskola:1998df} and Hirai et al.~\cite{Hirai:2001np}. However, some recent work has been done to discuss the nuclear 
medium effects in the cross sections and polarization of $\tau$ leptons produced in the $\nu_\tau$ induced quasielastic scattering~\cite{Graczyk:2004uy,Sobczyk:2019urm} but not
in the case of deep inelastic scattering induced by $\nu_\tau$. This is in contrast to the study of the deep inelastic scattering induced by the electron neutrinos 
($\nu_e$) and muon neutrinos ($\nu_\mu$) from nuclei in which the nuclear medium effects have 
been studied extensively~\cite{Haider:2011qs, SajjadAthar:2007bz, SajjadAthar:2009cr, Haider:2012nf, Haider:2012ic, Haider:2015vea, Haider:2016tev, Haider:2016zrk, Zaidi:2019mfd, Zaidi:2019asc, Kulagin:2004ie, Kulagin:2007ju}. 
Our aim in this paper is to study the nuclear medium effects in the deep inelastic scattering of tau neutrinos off nucleus in general and apply it to
the $^{40}Ar$ nucleus in the energy region relevant for the DUNE experiment.

The $\nu_\tau-$nucleon scattering from the free nucleons in the DIS region has been studied by many authors
~\cite{Albright:1974ts,Dasgupta:1996hh,Stein:1998wr,Kretzer:2003iu,Paschos:2001np,Jeong:2010nt,Hagiwara:2003di,Conrad:2010mh}.
The new features which appear in the case of the $\nu_\tau-$nucleon interaction as compared to the $\nu_e-$nucleon and $\nu_\mu-$nucleon interactions and 
contribute to modify the cross sections are:
\begin{itemize}
 \item the kinematical change in $Q^2$ and $E_l$ due to the presence of $m_\tau$, the finite mass of the $\tau$ lepton.
 \item the contributions due to the additional nucleon structure functions 
 $F_{4N}(x,Q^2)$ and $F_{5N}(x,Q^2)$ in the presence of $m_\tau\ne 0$.
 \item the modifications in cross sections due to the effect of polarization state of the $\tau$ leptons produced in the final state.
 \item the additional effects in the $Q^2$ evolution of the nucleon structure functions $F_{iN}(x,Q^2);~(i=1-5)$ due to $m_\tau \ne 0$ in the presence of massive heavy flavored quarks like the charm quark.
 \item the additional effects of the higher twist (HT)~\cite{Dasgupta:1996hh,Stein:1998wr} and the target mass corrections (TMC)~\cite{Kretzer:2003iu} on the structure functions $F_{iN}(x,Q^2);~(i=1-5)$ in the presence of $m_\tau \ne 0$ and $m_q \ne 0$.
\end{itemize}

Some of the above effects are modified in the nuclear medium and need to be calculated using a reliable nuclear model to describe the deep inelastic scattering of leptons from the nuclear targets. For example,
\begin{itemize}
 \item  the structure functions are modified due to the nuclear medium effects(NMEs). This was for the first time observed in the case of $F_{2A}(x,Q^2)$ and $F_{1A}(x,Q^2)$ by the EMC collaboration 
 and later on confirmed by many other experiments done with electrons and neutrinos.
 \item in the presence of nuclear medium effects, the nuclear structure functions  $F_{1A}(x,Q^2)$ and $F_{2A}(x,Q^2)$ may deviate from the Callan-Gross 
 relation~\cite{Callan:1969uq}, and the nuclear structure functions  $F_{4A}(x,Q^2)$ and $F_{5A}(x,Q^2)$ may not satisfy the Albright-Jarlskog relation~\cite{Albright:1974ts}. It is required that independently these nuclear structure functions are studied.
 \item the produced $\tau$ leptons in the final state may get depolarized in the nuclear medium affecting the production cross section from nuclear targets. The depolarization of $\tau$ will also affect the 
 topologies and characteristics of its decay products.
 \item there would be additional contributions to the structure functions due to non-nucleonic degrees of freedom in nucleilike pion and rho meson, except for $F_{3A}(x,Q^2)$ where only valence quarks contribute.
 \item the shadowing and the antishadowing effects in the respective kinematic regions of the Bjorken variable $x$ which are known to be present in the $\nu_\mu$-nucleus deep inelastic scattering 
 will also be present in the case of $\nu_\tau$-nucleus scattering and need to be taken into account.
\end{itemize}
In this work we report on the study of the deep inelastic scattering cross sections for the $\nu_\tau/\bar\nu_{\tau}-^{40}Ar$ scattering in the energy region relevant for the DUNE and atmospheric 
neutrino oscillation experiments.
The study includes nuclear medium effects mentioned above on the nucleon structure functions and cross sections except for the effect of the
depolarization of the $\tau$ lepton in the final state which is presently under investigation. The corrections due to the nuclear medium effects such as  the Fermi motion,  the binding energy and  
the nucleon correlations have been calculated using the spectral function~\cite{FernandezdeCordoba:1991wf} of the nucleons in the nucleus. The mesonic 
contribution is also calculated and is found to be significant in the low and intermediate region of  $x$ and  is incorporated following Ref.~\cite{Marco:1995vb,GarciaRecio:1994cn}. 
The (anti)shadowing corrections have been incorporated following the works of Kulagin and Petti~\cite{Kulagin:2004ie}. Furthermore, the effects of applying a cut of the center of 
mass energy ($W$) on the scattering cross section are also discussed. 

Since the nucleon structure functions are the basic inputs in the determination of nuclear structure functions and the scattering cross section, therefore, a proper understanding of 
the nucleon structure functions becomes quite important. In the low and moderate $Q^2$ region, the perturbative effects such as  the $Q^2$ evolution of the parton distribution functions from the leading order 
to the next-to-leading order (NLO), 
next-to-next-to-leading order (NNLO) as well as the nonperturbative effects like the kinematical higher twist effect that is also known as the 
target mass correction (TMC) which arises due to the massive quark contribution (e.g. charm, bottom, top) and dynamical higher twist effect (HT) which arises due to the multiparton correlations, become important. These nonperturbative effects
are important in the kinematical region of high $x$ and low $Q^2$~\cite{Ansari:2020xne}. 

\begin{figure}
 \includegraphics[height=8 cm, width=.9\textwidth]{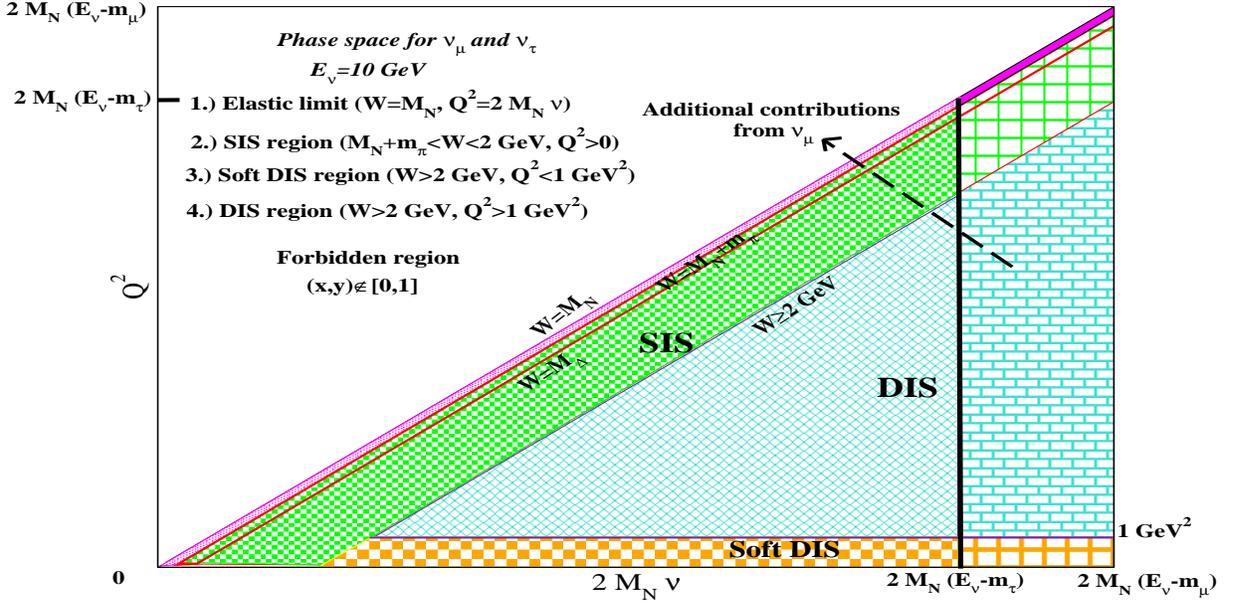}
\caption{$Q^2-\nu$ plane showing the allowed kinematic regions for $\nu_\mu$ and $\nu_\tau$ induced processes at $E_\nu=10$ GeV. The forbidden region 
is defined as $x,y~\notin~[0,1]$. For the elastic scattering, the CM energy is $W=M_N$ and Bjorken variable is
 $x=\frac{Q^2}{2M_N\nu}=1$. The SIS region has been defined as the
region for which $M_N+m_\pi \le W \le 2$ GeV and $Q^2 \ge 0$ covering both non-resonant and resonant meson production. The DIS region is defined as the region for 
which $Q^2 \ge 1$ GeV$^2$ and $W \ge 2$ GeV, and 
Soft DIS region is defined as $Q^2 < 1$ GeV$^2$ and $W \ge 2$ GeV. 
For the region lying in the band of $M_N<W<M_N+m_\pi$, we expect process like single photon emission. However, this region becomes important when the scattering takes place 
with a nucleon within a nucleus due to the multi-nucleon correlation effect. Soft DIS region is nothing but the SIS region. The band after the vertical black solid line is
depictively the additional contribution for each process like SIS, DIS, etc. to the scattering cross section.
The boundaries between regions are not sharply established and are indicative only.}
\label{fig_kin}
 \end{figure}

The inclusive cross sections at high energies and $Q^2$ , are expressed in terms of the structure functions corresponding to the deep 
inelastic scattering processes from the quarks and gluons. As one moves towards low energies one encounters the region of shallow inelastic scattering 
(SIS) which constitutes of the resonant and the nonresonant processes, with hadronic degrees of freedom. Presently, there is no sharp kinematic boundary to 
distinguish these two regions. In the SIS region, several resonances contribute to the scattering cross section and the nucleon to resonance transition
is described in terms of nucleon-to-resonance transition form factors. Presently these transition form factors are studied only for the 
$N\rightarrow \Delta(1232)$ and $N \rightarrow N^\ast(1440)$ transitions. All the (anti)neutrino experiments are being performed using nuclear targets and the properties 
of the resonances like their widths and masses may be modified in the nuclear medium, while there is not much study on the nuclear medium modifications on the properties of these resonances 
 except for the $\Delta(1232)$ resonance. Due to the absence of any sharp cut on the kinematical variables  defining the separation of the SIS and DIS regions, in literature, there is large variation
in the consideration of the values of $W$ and $Q^2$ on the onset of the 
DIS region. Lalakulich et al.~\cite{Lalakulich:2006yn}  have suggested a constrain of 1.1 GeV $\le W \le 1.6$ GeV on the center of mass energy in order to avoid the double counting of events in 
the transition region, while  Hagiwara et al.~\cite{Hagiwara:2003di} considers this limit to be 1.4 GeV $\le W \le 1.6$ GeV, whereas $W> 1.4$ GeV have been considered by Gazizov et al.~\cite{Gazizov:2016dhn}
and Kretzer et al.~\cite{Kretzer:2002fr}, as the onset of DIS processes. Besides theoretical studies, in the Monte Carlo event generators like NEUT~\cite{Hayato:2021heg} and GENIE~\cite{Andreopoulos:2015wxa} these 
boundaries are taken to be $W > 2$ GeV and $W > 1.7$ GeV, 
respectively, for the simulation of neutrino events.
The region of $W \ge 2$ GeV and $Q^2 \ge 1$ GeV$^2$ is considered to be the region of safe DIS or true DIS in MINERvA experiment~\cite{MINERvA:2021owq, MINERvA:2016oql}.
Recently this ambiguity in defining the onset of the DIS region has been discussed in the literature~\cite{SajjadAthar:2020nvy, Athar:2020kqn}.
In Fig.~\ref{fig_kin}, kinematic regions for the different processes such as elastic, inelastic, deep inelastic 
as well as soft DIS induced by $\nu_\mu$ and $\nu_\tau$ are shown. From the figure, one may notice the 
reduced kinematic region for $\nu_\tau$ events as compared to the allowed $\nu_\mu$ events. This is due to the
mass of the $\tau-$lepton which is $ \sim 17$ times heavier than the muon mass.
 
The proposed experiment, DUNE at the Fermilab is very promising and plans to resolve many subtle issues like a comprehensive investigation of neutrino oscillations to test CP violation in the lepton sector, determining the ordering of neutrino masses, etc. Due to the relatively broad and 
high energy neutrino spectrum at DUNE, about $40-50\%$ of the neutrino interactions 
will come from deep inelastic scattering rather than the quasielastic scattering and single pion production reactions ($\sim 40\%$ combined), and it
is expected that about 60\% of the events would come from the combined region of SIS+DIS. Therefore, it is 
important to understand the effect of the kinematical cut on the CM energy $W$ and $Q^2$ on the cross section, while evaluating the contribution of the DIS cross section to the total cross section. 
Therefore, in the present work, we have also studied the effect of the CM energy cut of 1.6 GeV
and 2 GeV keeping $Q^2 \ge 1$ GeV$^2$ on the evaluation of the nuclear structure functions and the differential cross sections.

In the present work, the nucleon structure functions have been evaluated using the MMHT PDFs parameterization~\cite{Harland-Lang:2014zoa} up to NLO in the four flavor 
($u~ d,~ s,$ and $c$) minimal subtraction (MSbar) scheme~\cite{Kretzer:2003iu}.  The nonperturbative effects of TMC and HT have been included following Refs.~\cite{Dasgupta:1996hh,Stein:1998wr} and 
~\cite{Kretzer:2003iu} respectively.
The QCD corrections have been first evaluated at the free nucleon level and then
the nuclear structure functions have been evaluated including the nuclear medium effects. 

This paper proceeds as follows: Section II presents a brief formalism for the (anti)neutrino- nucleus DIS process. This is followed by the discussion of the method for obtaining nuclear structure 
functions with nuclear medium effects due to the Fermi motion, the binding energy, the nucleon correlations, 
 the mesonic contribution and the (anti)shadowing. Section III presents the numerical results and their discussion. Section IV describes the summary of our findings.

 \section{Formalism}\label{sec_formalism}
  For the evaluation of the weak nuclear structure functions not much theoretical efforts have been made except that of Kulagin et al.~\cite{Kulagin:2004ie,Kulagin:2007ju} and Athar et al.
  (Aligarh-Valencia group)~\cite{SajjadAthar:2007bz, SajjadAthar:2009cr, Haider:2011qs, Haider:2012nf, Haider:2012ic, Haider:2015vea, Haider:2016tev, Haider:2016zrk,Zaidi:2019mfd, Zaidi:2019asc}.
    Aligarh-Valencia group has studied nuclear medium effects in the structure functions in a microscopic model 
 which uses relativistic nucleon spectral function to describe the target nucleon momentum distribution incorporating the effects of Fermi motion, binding energy and nucleon correlations in a field theoretical model. 
 The spectral function that describes the energy and momentum distribution of the nucleons in nuclei is
obtained by using the Lehmann's representation for the relativistic nucleon propagator and nuclear many body theory is used to calculate it for an interacting Fermi sea in the nuclear 
medium~\cite{FernandezdeCordoba:1991wf}. A local density approximation is then applied to translate these results to a finite nucleus. Furthermore, the 
contributions of the pion and rho meson clouds in a many body field theoretical approach have also been considered which is based on Refs.~\cite{Marco:1995vb,GarciaRecio:1994cn}. In this section, 
the theoretical approach of Aligarh-Valencia group is discussed briefly.

\begin{figure}
\begin{center}
\includegraphics[height=5 cm, width=7 cm]{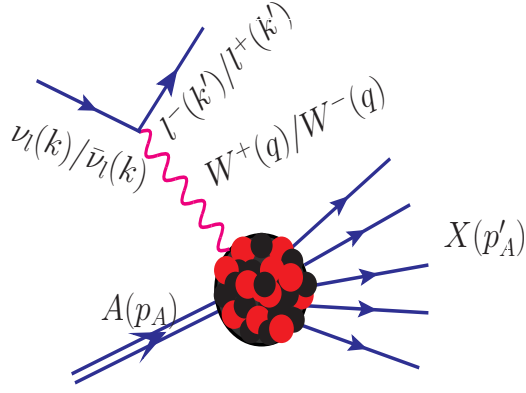}
\end{center}
\caption{Feynman diagrams for the $\nu_l/\bar\nu_l;~(l=\mu,\tau)$ induced DIS process off nuclear target ($A$).}
\label{dis_nuc}
\end{figure}

 The differential scattering cross section for the charged current inclusive $\nu_l/\bar\nu_l$-nucleus deep inelastic scattering process (depicted in Fig.~\ref{dis_nuc}):
\begin{equation}\label{DISrxn}
 \nu_l / \bar\nu_l(k) + A(p_A) \rightarrow l^-/l^+(k') + X(p'_A)
\end{equation}
 is expressed in terms of the leptonic tensor $L_{\mu\nu}$ and 
  the nuclear hadronic tensor $W^{\mu\nu}_A$ as:
\begin{eqnarray}\label{xecA}
{ d^2\sigma_A \over dx dy }&=& \left({G_F^2 y M_N E_l  \over 2\pi E_{\nu}}\right) {\left(M_W^2\over M_W^2+Q^2 \right)^2} \; {|{\bf k^\prime}| \over|{ \bf k}|}\;
L_{\mu\nu}\; W^{\mu\nu}_A,
\end{eqnarray}
where in Eq.\ref{DISrxn}, the quantities in the brackets are the four momenta of the corresponding particles, for example, $k$ is the 
four momentum of incoming neutrino, $p_A$ is the four momentum of the initial target nucleus, and so on. In Eq.\ref{xecA}, $G_F$ is the Fermi coupling constant, $M_N$ is the nucleon mass, $E_\nu$ and 
$E_l$ are respectively, the energies of the incoming neutrino and the outgoing charged lepton. $M_W$ is the mass of intermediate $W-$boson propagator, $Q^2$ is the four momentum transfer 
square, $x$ is the Bjorken variable and $y$ is the inelasticity. The leptonic tensor $L_{\mu \nu}$ is given by
\begin{eqnarray}\label{lep_weak}
L_{\mu \nu} &=&8(k_{\mu}k'_{\nu}+k_{\nu}k'_{\mu}
-k.k^\prime g_{\mu \nu}  \pm i \epsilon_{\mu \nu \rho \sigma} k^{\rho} 
k'^{\sigma})\,,
\end{eqnarray}
where $+/-$ sign is for $\nu_l/\bar \nu_l$. The nuclear hadronic tensor $W^{\mu\nu}_A$ is written in terms of the weak nuclear structure functions $W_{iA}(\nu_A,Q^2);$ ($i=1-6$) as:
\begin{eqnarray}
 \label{nuc_had_weak}
W_{A}^{\mu \nu} &=&
\left( \frac{q^{\mu} q^{\nu}}{q^2} - g^{\mu \nu} \right) \;
W_{1A} (\nu_A, Q^2)
+ \frac{W_{2A} (\nu_A, Q^2)}{M_A^2}\left( p^{\mu}_A - \frac{p_A . q}{q^2} \; q^{\mu} \right)\left( p^{\nu}_A - \frac{p_A . q}{q^2} \; q^{\nu} \right)
 \nonumber\\
&\pm&\frac{i}{2M_A^2} \epsilon^{\mu \nu \rho \sigma} p_{A \rho} q_{\sigma}
W_{3A} (\nu_A, Q^2) + \frac{W_{4A} (\nu_A, Q^2)}{M_A^2} q^{\mu} q^{\nu}+\frac{W_{5A} (\nu_A, Q^2)}{M_A^2} (p^{\mu}_A q^{\nu} + q^{\mu} p^{\nu}_A)\nonumber\\
&+& \frac{i}{M_A^2} (p^{\mu}_A q^{\nu} - q^{\mu} p^{\nu}_A)
W_{6A} (\nu_A, Q^2)\,,
\end{eqnarray}
where $M_A$ is the mass of the nuclear target. $W_{6A}(\nu_A,Q^2)$ does not contribute to the cross section as it vanishes when contracted with the leptonic tensor $L_{\mu \nu}$.
 The nuclear structure functions $W_{iA}(\nu_A,Q^2)~(i=1-5)$ are written in terms of the dimensionless nuclear structure functions $F_{iA}(x_A);~(i=1-5)$ as~\cite{Zaidi:2019asc, Kretzer:2003iu}:
 
%
\begin{equation}\label{relation1}
\left.
\begin{array}{l}
 F_{1A}(x_A) =M_A W_{1A}(\nu_A,Q^2) \\
 F_{2A}(x_A) = \frac{Q^2}{2xM_A}W_{2A}(\nu_A,Q^2)\\
 F_{3A}(x_A) = \frac{Q^2}{xM_A}W_{3A}(\nu_A,Q^2)\\
 F_{4A}(x_A) = \frac{Q^2}{2M_A}W_{4A}(\nu_A,Q^2) \\
 F_{5A}(x_A) = \frac{Q^2}{2xM_A}W_{5A}(\nu_A,Q^2), 
 \end{array}
 \right\}
\end{equation}
where $\nu_A$(=$\frac{p_{_A}\cdot q}{M_{_A}}(=q^{0})$ ) is the energy transferred to the nuclear target in the rest frame of the nucleus i.e. $p_A=(p_A^0,~{\bf p_A}= 0)$
and $x_A(=\frac{Q^2}{2 p_A \cdot q}=\frac{Q^2}{2 p_{A}^0  q^0 } = \frac{Q^2}{2 A~M_N q^0}=\frac{x}{A})$ is the Bjorken scaling variable corresponding to the nucleus.

The expression for the differential cross section for the $\nu_l/{\bar\nu}_l - A$ scattering can be obtained using Eqs.~\ref{lep_weak},  \ref{nuc_had_weak} and \ref{relation1} in Eq.~\ref{xecA} as
\begin{eqnarray}\label{xsecsf}
 \frac{d^2\sigma_A}{dxdy}&=&\frac{G_F^2M_NE_\nu}{\pi(1+\frac{Q^2}{M_W^2})^2}
 \Big\{\Big[y^2x+\frac{m_l^2 y}{2E_\nu M_N}\Big]F_{1A}(x,Q^2)+
 \Big[\Big(1-\frac{m_l^2}{4E_\nu^2}\Big)-\Big(1+\frac{M_Nx}{2E_\nu}\Big)y\Big]F_{2A}(x,Q^2)\nonumber\\
 &\pm& \Big[xy\Big(1-\frac{y}{2}\Big)-
 \frac{m_l^2 y}{4E_\nu M_N}\Big]F_{3A}(x,Q^2)
 +\frac{m_l^2(m_l^2+Q^2)}{4E_\nu^2M_N^2 x}F_{4A}(x,Q^2)-\frac{m_l^2}{E_\nu M_N}F_{5A}(x,Q^2)\Big\}.\;\;\;
\end{eqnarray}
 The scaling variables $x\big(=\frac{Q^2}{2 p \cdot q}\big)$ and $y(=\frac{\nu}{E_\nu}=\frac{q_0}{E_\nu})$ lie in the range:
\begin{equation}\label{xydef}
 \frac{m_l^2}{2M_N (E_\nu - m_l)} \le x \le 1~~~~~~~~~\text{and}~~~~~~~~~~~
 a-b \le y\le a+b,
\end{equation}
where
\begin{eqnarray}
 a=\frac{1-m_l^2\Big(\frac{1}{2M_NE_\nu x}+\frac{1}{2E_\nu^2} \Big)}{2\Big(1+\frac{M_N x}{2E_\nu}\Big)}~~~~~~~~\text{and}~~~~~~~~
 b=\frac{\sqrt{\left(1-\frac{m_l^2}{2 M_N E_\nu x}\right)^2-\frac{m_l^2}{E_\nu^2}}}{2\Big(1+\frac{M_N x}{2E_\nu}\Big)}.
\end{eqnarray} 
For $\nu_e/\bar{\nu}_e$ and $\nu_\mu/\bar{\nu}_\mu$ interactions with a nuclear target (i.e. in the limit $m_l\to 0$), only the first three terms of Eq.~\ref{xsecsf}, i.e. the terms with
$F_{1A}(x,Q^2)$, $F_{2A}(x,Q^2)$ and $F_{3A}(x,Q^2)$ would contribute. However, for $\nu_\tau/\bar{\nu}_\tau$ all the five structure functions ($F_{iA}(x,Q^2);~(i=1-5)$) would contribute as the
terms with tau-lepton mass ($m_\tau=1.78$ GeV) can not be ignored.
In the laboratory frame, the nuclear target is at rest but the nucleons bound inside the nucleus are moving continuously with a finite momentum, i.e. ${\bf p_N}$ is non-zero
and the motion of such nucleons corresponds to the Fermi motion.
 If the momentum transfer is along the $Z$-axis then $q^\mu=(q^0,0,0,q^z)$ and the Bjorken variable $x_N$ corresponding to the nucleon bound inside a nucleus is written as:
 \begin{equation}
 x_N = \frac{Q^2}{2 p_N \cdot q} = \frac{Q^2}{2 (p_N^0 q^0 - p_N^z q^z)}.
\end{equation}
The bound nucleons interact with each other through the strong force hence various nuclear medium effects come into the picture. Depending upon the value of the Bjorken variable $x$ the various nuclear medium effects have different contribution.
  The nuclear medium effects such as Fermi motion, binding, nucleon correlations, meson cloud contribution and shadowing effect are discussed in the Subsections II.1, II.2 and II.3, respectively.

\subsubsection{Fermi motion, binding and nucleon correlation}\label{spec}
In order to calculate the cross section for the neutrino scattering off a bound nucleon inside the nucleus in the presence of nuclear medium, we begin with a neutrino flux hitting  the target 
nucleons over a given period of time. Since neutrinos are the weakly interacting particles; therefore  majority of them will 
pass through the target without having any interaction while a few neutrinos will
 interact with the target nucleons giving rise to final state leptons and
 hadrons.  To consider the interaction of neutrinos, we introduce the concept of ``neutrino self-energy". The real part of ``neutrino self-energy" modifies the lepton mass and imaginary part 
 gives information about the total number of neutrinos interaction that yield the final state leptons and hadrons.

 The cross section ($d\sigma_A$) for small elemental volume ($dV$) inside the nucleus is 
related to  the probability of neutrino interaction with a bound nucleon per unit time ($\Gamma$). Probability times the differential of area ($dS$) defines the cross section ~\cite{Marco:1995vb}, i.e.
\begin{equation}\label{defxsec}
d^2\sigma_A=\Gamma dt dS = \Gamma \frac {E_\nu({\bf k})}{\mid   {\bf k}  \mid}d^3 r,~~~~~~~~~\Big[\because~~dtdS= \frac{dV}{v}=\frac {E_\nu({\bf k})}{\mid {\bf k}\mid }d^3 r \Big]
\end{equation}
where $v$ is the velocity of the incoming neutrino. $\Gamma$ is related to the imaginary part of $\nu_l$ self-energy ($\Sigma(k)$) as~\cite{Marco:1995vb}:
\begin{equation}\label{eqr}
-\frac{\Gamma}{2} = \frac{m_\nu}{E_\nu({\bf k})}\; Im \Sigma(k).
\end{equation}
From Eq.\ref{defxsec} and Eq.\ref{eqr}, we get
\begin{equation}\label{eqq}
 d^2\sigma_A= -2\frac{m_\nu}{\mid {\bf k} \mid} Im \Sigma (k) d^3 r.
\end{equation}

\begin{figure}
\begin{center}
 \includegraphics[height=4.0 cm, width=11 cm]{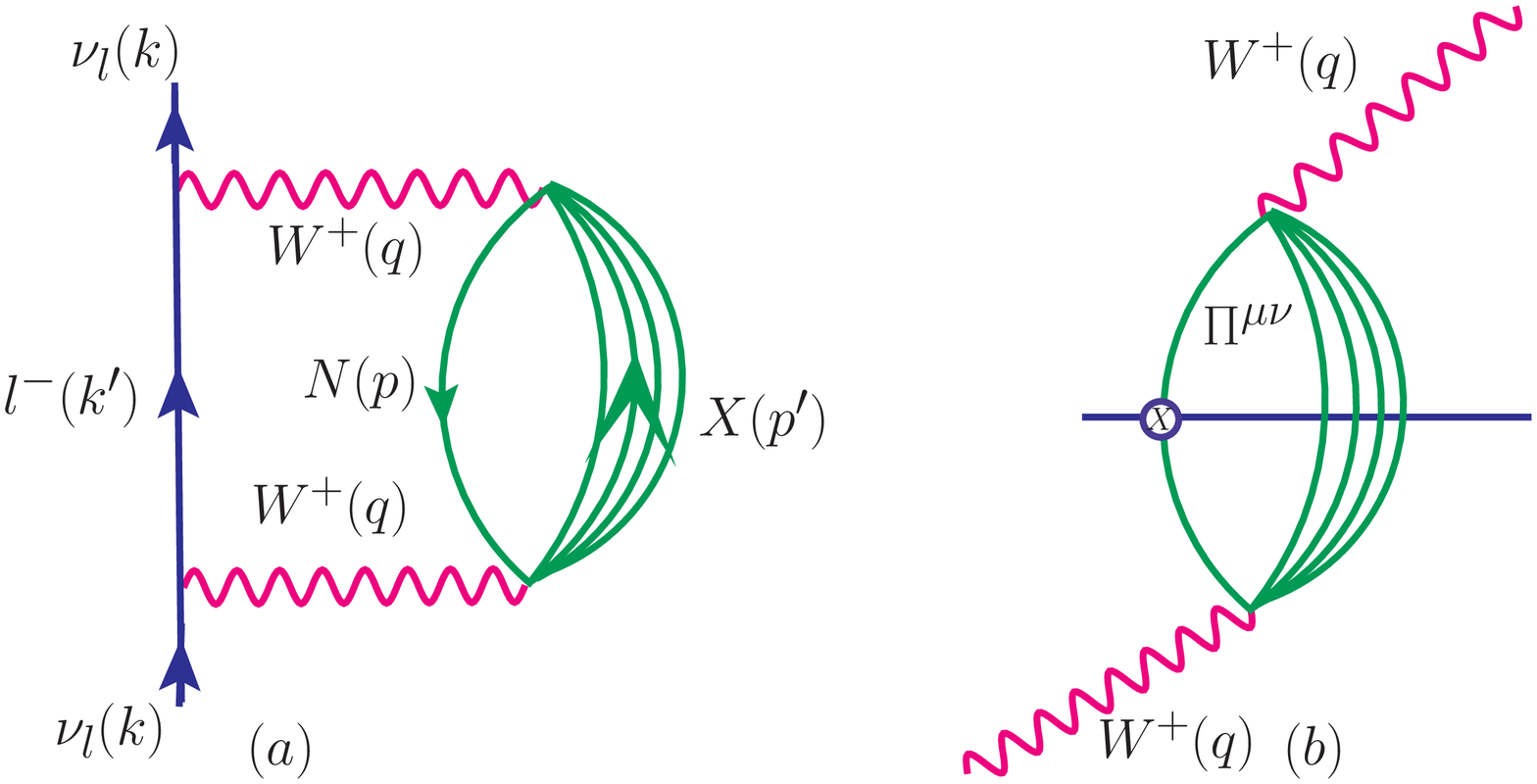}
 \end{center}
 \caption{Diagrammatic representation of {\bf (a)} the neutrino self-energy and {\bf (b)} the intermediate vector boson $W^+$ self-energy.}
 \label{wself_energy}
\end{figure}
\begin{figure}[h]
\begin{center}
 \includegraphics[height=2.5 cm, width=12 cm]{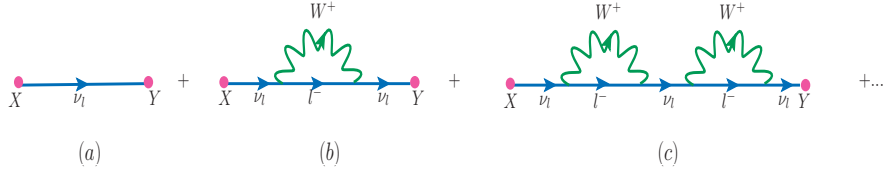}
 \end{center}
 \caption{Fermion two point function and its modification.}
 \label{self_fig}
\end{figure}

 The neutrino self-energy is evaluated corresponding to the diagram shown in Fig.\ref{wself_energy} (left panel). In  many body field theory the interaction of neutrino with a potential provided by a nucleus
can be explained as the modification to the fermion two point function as depicted  in Fig.\ref{self_fig}.

Figure \ref{self_fig}(a) corresponds to the free field fermion propagator while Figure \ref{self_fig}(b,c) constitutes to the neutrino self-energy. 
Using the Feynman rules we write the neutrino self-energy corresponding to Figure \ref{wself_energy}(a) as
\begin{eqnarray}
    -i \Sigma(k)&=&\int \frac{d^4 q}{(2 \pi)^4}\Big( \bar u_\nu(k)\frac{-ig}{2\sqrt{2}}\gamma_\mu(1-\gamma_5)\times \frac{i(\not k'+m_l)}{k^{'2}-m_l^2+i\epsilon} \frac{-ig}{2\sqrt{2}}\gamma_\nu(1-\gamma_5)u_\nu(k)\Big)\nonumber\\
    &\times& \Big(-\frac{i g^{\mu\rho}}{q^2-M_W^2} \Big)(-i\Pi_{\rho\sigma}(q))\Big(-\frac{i g^{\sigma\nu}}{q^2-M_W^2} \Big).
\end{eqnarray}
Now we use the relations 
\begin{eqnarray}
\frac{g^2}{8 M_W^2} &=& \frac{G_F}{\sqrt{2}};\hspace{4 mm}
d^4 q=d^4 k';\hspace{4 mm}
\sum_r u_r(k) \bar u_r(k)=\frac{\not k+m_\nu}{2 m_\nu}\nonumber
\end{eqnarray}
and also apply the Cutkowsky rules
 \begin{eqnarray}
\Sigma(k) &\rightarrow& 2 i Im \Sigma(k);~~~\textrm{Lepton~ self-energy} \nonumber\\
\Pi^{\mu\nu}(q) &\rightarrow& 2 i \theta(q^0) Im \Pi^{\mu\nu}(q);~~~\textrm{W$^+$~boson~ self-energy},\nonumber
\end{eqnarray}
to obtain the imaginary part of the neutrino self-energy $Im \Sigma (k)$  as:
\begin{equation}\label{nu_imslf1}
Im \Sigma(k)=\frac{ G_F}{\sqrt{2}} {4 \over m_\nu} \int \frac{d^3 k^\prime}{(2 \pi)^4} {\pi \over E({\bf k^\prime})} \theta(q^0) \left(\frac{M_W}{Q^2+M_W^2}\right)^2\;Im[L_{\mu\nu}^{WI} \Pi^{\mu\nu}(q)],
\end{equation}
where $\Pi^{\mu\nu}(q)$ is the $W^+$-boson self-energy (as shown in Figure~\ref{wself_energy}(b)).

$\Pi^{\mu\nu}(q)$ is generally written in terms of the nucleon propagator ($G_l$)  and meson propagator ($D_j$) corresponding to  Figure~\ref{wself_energy}(b), as:
\begin{eqnarray}\label{wboson}
 \Pi^{\mu \nu} (q)&=& \left(\frac{G_F M_W^2}{\sqrt{2}}\right) \times \int \frac{d^4 p}{(2 \pi)^4} G (p) 
\sum_X \; \sum_{s_p, s_l} \prod_{i = 1}^{N} \int \frac{d^4 p'_i}{(2 \pi)^4} \; \prod_{_l} G_l (p'_l)\prod_{_j} \; D_j (p'_j)\; \nonumber \\  
&&  <X | J^{\mu} | N >  <X | J^{\nu} | N >^* (2 \pi)^4 ~
\delta^4 \Big(k + p - k^\prime - \sum^N_{i = 1} p'_i\Big),\;\;\;
\end{eqnarray}
where $s_p$ and $s_l$ are the spins of the initial state nucleon and the final state fermions, the indices $l$ and $j$ are respectively, for the fermions and bosons in the final hadronic state, $<X | J^{\mu} | N >$ represents the hadronic current and $\delta^4 (k + p - k^\prime - \sum^N_{i = 1} p'_i)$ ensures the conservation
of four momentum. $G(p)$ gives the information about the propagation of the nucleon from the initial state to the final state or vice versa. 

\begin{figure}
\begin{center}
 \includegraphics[height=5 cm, width=12.5 cm]{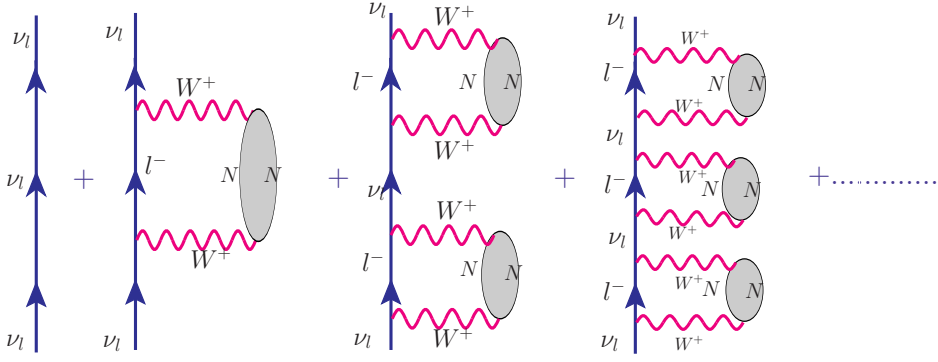}
 \end{center}
 \caption{Diagrammatic representation of neutrino self-energy in the nuclear medium.}
 \label{n_self}
\end{figure}

 To obtain the relativistic nucleon propagator $G(p^0,{\bf p})$ in the nuclear medium we start with 
 the relativistic free nucleon Dirac propagator $G^{0}(p^{0},{\bf p})$, which is written in terms of the Dirac spinors for particles $u({\bf p})$ and antiparticles $v({\bf p})$.
 This includes the contribution from positive and negative energy components of the nucleon, where the negative energy 
 contribution is suppressed while the positive energy contribution survives~\cite{Marco:1995vb,FernandezdeCordoba:1991wf}. Therefore, the free nucleon propagator may be expressed as
\begin{eqnarray}
 G^{0}(p^{0},{{\bf p}}) = \frac{1}{\not p - M_N+i \epsilon}=\frac{\not p+M_N}{(p^2-M_N^2+i\epsilon)}. 
\end{eqnarray}
 Considering only the positive energy part the above expression gets modified to
 \begin{equation}
 G^{0}(p^{0},{{\bf p}})= {\not p +M_N \over p^2-M_N^2 +i\epsilon} + 2\ i \pi \theta(p^0) \delta(p^2 -M_N^2) n({\bf p}) (\not p +M_N)
\end{equation}
 In the interacting Fermi sea, the relativistic nucleon propagator $G(p^0,{\bf p})$ is written
in terms of the 
nucleon self-energy $\Sigma^N(p^0,\bf{p})$ (depicted in Fig.\ref{n_self}), which contains all the information on single nucleon. Then in nuclear medium the interaction is taken into account through Dyson series expansion, which can be understood as 
the quantum field theoretical analogue of the Lippmann-Schwinger equation for the dressed nucleons, which is in principle an infinite series in perturbation theory. We add 
this perturbative expansion in a ladder approximation (Fig.\ref{n_self}) as: 
\begin{eqnarray}\label{gpseries}
 G(p) &=& G^0(p)~+~G^0(p)\Sigma^N(p)G^0(p)~+~G^0(p)\Sigma^N(p)G^0(p)\Sigma^N(p)G^0(p)~+~.......,\nonumber
\end{eqnarray}
which after simplification modifies to
\begin{eqnarray}\label{gp1}
G(p)&=&\frac{M_N}{E({\bf p})}\frac{\sum_{r}u_{r}({\bf p})\bar u_{r}({\bf p})}{p^{0}-E({\bf p})}+\frac{M_N}{E({\bf p})}\frac{\sum_{r}u_{r}({\bf p})\bar
u_{r}({\bf p})}{p^{0}-E({\bf p})}\Sigma^N(p^{0},{\bf p})
 \frac{M_N}{E({\bf p})} \frac{\sum_{s}u_{s}({\bf p})\bar u_{s}({\bf p})}{p^{0}-E(\bf p)}+..... \nonumber \\
&=&\frac{M_N}{E({\bf p})}\frac{\sum_{r} u_{r}({\bf p})\bar u_{r}({\bf p})}{p^{0}-E({\bf p})-\sum_{r}\bar u_{r}({\bf p})\Sigma^N (p^{0},{\bf p})u_{r}({\bf p})
\frac{M_N}{{E({\bf p})}}}\;\;\;
\end{eqnarray}
 The spin diagonal nucleon self-energy is written using spinorial indices
 $\alpha$ and $\beta$ as $\Sigma^N_{\alpha\beta}(p^{0},{\bf p})=\Sigma^N(p^{0},{\bf p})\delta_{\alpha\beta}$. $\Sigma^N(p)$ is taken 
 from Ref.~\cite{FernandezdeCordoba:1991wf, Oset:1981mk} and is obtained using the techniques of standard many body theory. Imaginary part of the nucleon self-energy is calculated explicitly and then $Re \Sigma^N(p^{0},{\bf p})$
 is obtained by means of dispersion relations using $Im \Sigma^N(p^{0},{\bf p})$. 

 In the nuclear matter the dressed nucleon propagator is written as~\cite{Marco:1995vb}:
 \begin{eqnarray}\label{Gp}
G (p) =&& \frac{M_N}{E({\bf p})} 
\sum_r u_r ({\bf p}) \bar{u}_r({\bf p})
\left[\int^{\mu}_{- \infty} d \omega 
\frac{S_h (\omega, {\bf{p}})}{p^0 - \omega - i \eta}
+ \int^{\infty}_{\mu} d  \omega 
\frac{S_p (\omega, {\bf{p}})}{p^0 - \omega + i \eta}\right],\;\;~~~
\end{eqnarray}
where the expression for the nucleon self-energy ($\Sigma^N(p^0,{\bf{p}})$) is taken from Ref.~\cite{FernandezdeCordoba:1991wf}. In the above 
expression $S_h (\omega, {\bf{p}})$, $S_p (\omega, {\bf{p}})$, $\mu(=\epsilon_F+M_N)$ and $\omega=p^0-M_N$ are the hole spectral function, particle spectral function, 
chemical potential and the removal energy, respectively. $\eta$ is the infinitesimal quantity i.e. $\eta \to 0$. In our earlier work~\cite{Haider:2015vea} the spectral 
function has been discussed in detail and for an inelastic scattering (Eq.\ref{DISrxn}) we need only the hole spectral function.

Then using Eqs.~\ref{eqq} and \ref{nu_imslf1} the expression for the differential cross section is written as 
\begin{equation}\label{dsigma_3}
\frac {d^2\sigma_A}{dx dy}=-\frac{G_F^2\;M_N\;y}{2\pi}\;\frac{E_l}{E_\nu}\;\frac{|\bf{k^\prime}|}{|\bf {k}|}\left(\frac{M_W^2}{Q^2+M_W^2}\right)^2 L_{\mu\nu} \int  Im \Pi^{\mu\nu}(q) d^{3}r.
\end{equation}
Comparing Eq.\ref{dsigma_3}, with Eqs.\ref{xecA}, \ref{wboson} and \ref{Gp}, the nuclear hadronic tensor (for isospin symmetric nucleus) can be expressed in terms of 
the nucleon hadronic tensor and the hole spectral function and is given as~\cite{Haider:2015vea}
\begin{equation}\label{conv_WAa}
W^{\mu \nu}_{A} = 4 \int \, d^3 r \, \int \frac{d^3 p}{(2 \pi)^3} \, 
\frac{M_N}{E ({\bf p})} \, \int^{\mu}_{- \infty} d p^0 S_h (p^0, {\bf p}, \rho(r))
W^{\mu \nu}_{N} (p, q), \,
\end{equation}
where $\rho(r)$ is the nucleon charge density inside the nucleus and  a factor of 4 is because of the spin-isospin degrees of freedom of the nucleon. 
For argon, we have used the two parameter Fermi density given by
\begin{equation}
 \rho(r)=\frac{\rho_0}{1+\exp(\frac{r-R}{a})},
\end{equation}
and the density parameters ($R=3.53fm$ and $a=0.542fm$) are taken from the electron-nucleus scattering experiments~\cite{DeVries:1987}.

  From Eq.~\ref{conv_WAa}, we see that the nuclear hadronic tensor $W^{\mu \nu}_{A}$ is written in terms of the nucleonic tensor $W^{\mu \nu}_{N}$ given by
 \begin{eqnarray}
 \label{nucleon_had_ten_weak}
W_{N}^{\mu \nu} &=&
\left( \frac{q^{\mu} q^{\nu}}{q^2} - g^{\mu \nu} \right) \;
W_{1N} (\nu_N, Q^2)
+ \frac{W_{2N} (\nu_N, Q^2)}{M_N^2}\left( p^{\mu}_N - \frac{p_N . q}{q^2} \; q^{\mu} \right)
 \nonumber\\
&\times&\left( p^{\nu}_N - \frac{p_N . q}{q^2} \; q^{\nu} \right)\pm\frac{i}{2M_N^2} \epsilon^{\mu \nu \rho \sigma} p_{N \rho} q_{\sigma}
W_{3N} (\nu_N, Q^2) + \frac{W_{4N} (\nu_N, Q^2)}{M_N^2} q^{\mu} q^{\nu}\nonumber\\
&&+\frac{W_{5N} (\nu_N, Q^2)}{M_N^2} (p^{\mu}_N q^{\nu} + q^{\mu} p^{\nu}_N)
+ \frac{i}{M_N^2} (p^{\mu}_N q^{\nu} - q^{\mu} p^{\nu}_N)
W_{6N} (\nu_N, Q^2)\,,
\end{eqnarray}
 where $W_{iN} (\nu_N, Q^2);~(i=1-6)$ are the nucleon structure functions, which in turn are expressed in terms of the dimensionless nucleon structure functions viz. $F_{iN}(x_N)~(i=1-5)$ as~\cite{Zaidi:2019asc, Kretzer:2003iu}:
 \begin{equation}\label{relation0}
\left.
\begin{array}{l}
 F_{1N}(x_N) = M_N W_{1N}(\nu_N,Q^2) \\
 F_{2N}(x_N) = \frac{Q^2}{2x_NM_N}W_{2N}(\nu_N,Q^2)\\
 F_{3N}(x_N) = \frac{Q^2}{x_NM_N}W_{3N}(\nu_N,Q^2)\\
 F_{4N}(x_N) = \frac{Q^2}{2M_N}W_{4N}(\nu_N,Q^2) \\
 F_{5N}(x_N) = \frac{Q^2}{2x_NM_N}W_{5N}(\nu_N,Q^2).
 \end{array}
 \right\}
\end{equation}
In the Bjorken limit, i.e. $Q^2\to \infty, ~\nu\to\infty$ with $x\to finite$, the dimensionless nucleon structure functions depend only on a single dimensionless variable $x$. However, if we move 
towards the region of low and moderate $Q^2$, these structure functions show $Q^2$ dependence and therefore become the functions of $x$ as well as $Q^2$. The dimensionless nucleon structure functions
are generally expressed in terms of parton distribution functions at the leading order, for example, 
\begin{eqnarray}
 &&F_{2N}^{\nu}(x)=x[u(x)+\bar u(x)+d(x)+\bar d(x)+2 s(x)+2 \bar c(x)];\hspace{3 mm}F_{2N}^{\bar\nu}=x[u(x)+\bar u(x)+d(x)+\bar d(x)+2 \bar s(x)+2 c(x)],\nonumber\\
 &&x F_{3N}^{\nu}(x)=x[u(x)-\bar u(x)+d(x)-\bar d(x)+2 s(x)-2 \bar c(x)];\hspace{3 mm}xF_{3N}^{\bar\nu}=x[u(x)-\bar u(x)+d(x)-\bar d(x)-2 \bar s(x)+2 c(x)],\nonumber\\
 && F_{4N}^{\nu/\bar\nu}(x)=0,
\end{eqnarray}
where $u(x)/\bar u(x)$ represents the probability density of finding an up quark/antiquark with a momentum fraction $x$. 
For $F_{1N}(x)$ and $F_{5N}(x)$, we have used the Callan-Gross relation ($F_{2N}(x)=2 x F_{1N}(x)$) and Albright-Jarlskog relation ($F_{2N}(x)=2x F_{5N}(x)$) at the leading order. One may notice that at the leading order $F_{4N}(x)=0$ but when the contribution
from the next-to-leading order terms is taken into account, we find that $F_{4N}(x)$ gives a non-zero contribution.
To evaluate the weak dimensionless nuclear structure functions by using Eq.\ref{conv_WAa}, the appropriate components of the nucleon ($W^{\mu\nu}_N$ in Eq.\ref{nucleon_had_ten_weak}) and the nuclear ($W^{\mu\nu}_A$ in Eq.\ref{nuc_had_weak}) hadronic tensors along 
the $x$, $y$ and $z$ axis are chosen. For example, the expression of nuclear structure function $F_{1A,N}(x_A,Q^2)$ incorporating the nuclear medium effects like binding energy, Fermi motion 
and nucleon correlations is obtained by taking the $xx$ components, $F_{3A,N}(x_A,Q^2)$ by taking the $xy$ components, etc. We obtain the expressions for all the five nuclear structure functions as: 
  \begin{eqnarray}\label{spect_funct}
F_{iA,N} (x_A, Q^2) &=& 4\int \, d^3 r \, \int \frac{d^3 p}{(2 \pi)^3} \, 
\frac{M_N}{E_N ({\bf p})} \, \int^{\mu}_{- \infty} d p^0~ S_h(p^0, {\bf p}, \rho(r))~
\times f_{iN}(x,Q^2)
\end{eqnarray}
where $i=1-5$ and
\begin{eqnarray}
 f_{1N}(x,Q^2)&=&AM_N\left[\frac{F_{1N} (x_N, Q^2)}{M_N} + \left(\frac{p^x}{M_N}\right)^2 \frac{F_{2N} (x_N, Q^2)}{\nu_N}\right],\\
f_{2N}(x,Q^2)&=&\left( \frac{F_{2N}(x_N,Q^2)}{M_N^2 \nu_N}\right)\left[ \frac{Q^4}{q^0 {(q^z)}^2}\left(p^z+\frac{q^0(p^0-\gamma p^z) }{Q^2}{ q^z} \right)^2+\frac{q^0 Q^2 (p^x)^2}{{(q^z)}^2}\right]\\
f_{3N}(x,Q^2)&=&A\frac{q^0}{q^z} \;\times\left({p^0 q^z - p^z q^0  \over p \cdot q} \right)F_{3N} (x_N,Q^2),\\
f_{4N}(x,Q^2)&=&A\; \left[F_{4N}(x_N,Q^2) +\frac{p^z Q^2}{{ q^z}} \frac{F_{5N}(x,Q^2)}{M_N \nu_N}\right],\\
f_{5N}(x,Q^2)&=&A\;\;\frac{F_{5N}(x_N,Q^2)}{M_N \nu_N}\times\left[q^0(p^0-\gamma p^z)+Q^2 \frac{p^z}{{ q^z}} \right]
\end{eqnarray}
 The nonperturbative effects of the target mass correction and the higher twist~\cite{Dasgupta:1996hh, Stein:1998wr} have been incorporated  in the free nucleon structure functions and then we have convoluted these 
nucleon structure functions with the spectral function in order to evaluate the nuclear structure
functions (Eq.\ref{spect_funct}). Using the nuclear structure functions, we have obtained the differential scattering cross 
sections  for the $\nu_\tau(\bar\nu_\tau)-A$ DIS process (Eq.\ref{xsecsf}). 

The calculations are performed in the four flavor MS-bar scheme, the light quarks $u,~d$ and $s$ are treated to be massless and charm quark $c$ to be a massive object. Hence, we define 
\begin{equation}\label{nf4_charm}
 F_{iA}(x,Q^2)=F_{iA}^{n_f=4}(x,Q^2)=\underbrace{F_{iA}^{n_f=3}(x,Q^2)}_{\textrm{for massless($u,d,s$) quarks}}+\underbrace{F_{iA}^{n_f=1}(x,Q^2)}_{\textrm{for massive charm quark}}.
\end{equation}
It is important to point out that in the case of massive charm contribution, $F_{iA}^{n_f=1}(x,Q^2);~(i=1-5)$ are target mass corrected~\cite{Kretzer:2003iu}, however, HT effect has not been 
included as there is no explicit prescription available in the literature to include this effect. 
\begin{figure}
\begin{center}
\includegraphics[height=4. cm, width=3. cm]{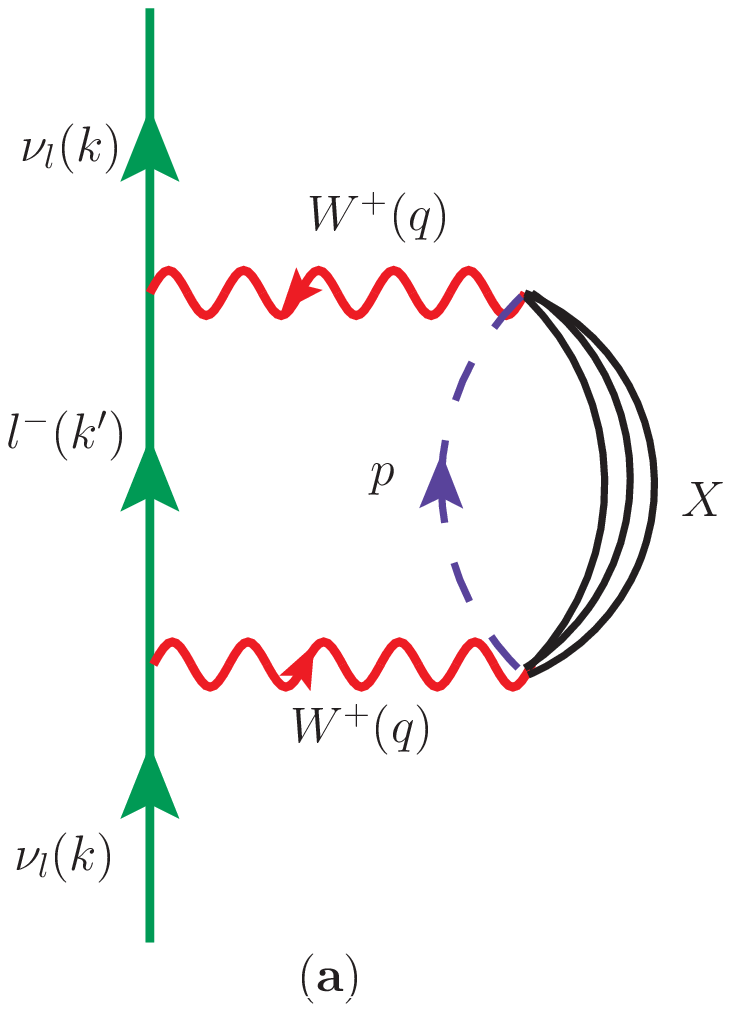}
 \includegraphics[height=4. cm, width=9 cm]{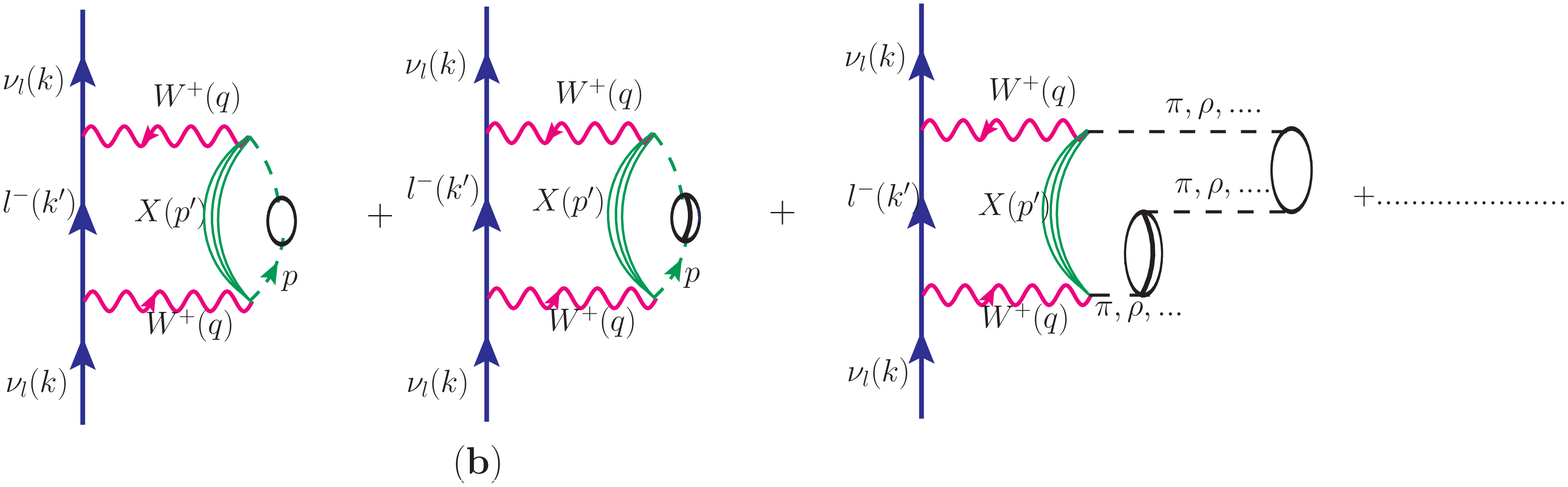}
 \end{center}
 \caption{Neutrino self-energy diagram accounting for neutrino-meson DIS (a) the bound nucleon propagator is substituted with a meson($\pi$ or $\rho$) propagator with momentum $p$
 represented here by dashed line
 (b) by including particle-hole $(1p–1h)$, delta-hole $(1\Delta–1h)$,
 $1p1h-1\Delta1h$, etc. interactions.}
 \label{n_self-1}
\end{figure}

The nucleons which are bound inside the nucleus may interact with each other via meson exchange such as 
$\pi,~\rho,$ etc., and the interaction of the intermediate vector boson (IVB) 
 with the mesons play an important role in the evaluation of nuclear structure functions~\cite{Haider:2012nf, Haider:2016zrk}. 
 This mesonic effect has been incorporated and is discussed in the following section~(\ref{mes}).

 \subsubsection{Contribution of pion and rho meson to the nuclear structure function (Mesonic effect)}\label{mes}
 Associated with each nucleon bound inside the nucleus there are virtual mesons (pion, rho meson, etc.) and because of the strong
 attractive nature of the nucleon-nucleon interaction, the probability of a $W^\pm$-boson interaction with the mesonic cloud becomes high.  
 In this work we have included the $\pi$ and $\rho$ meson contributions~\cite{Marco:1995vb, Kulagin:2004ie, Ericson:1983um, LlewellynSmith:1983vzz} as the contribution from heavier mesons are expected to be
 very small due to their significantly higher masses.
 The pion cloud contribution is larger than that of rho-meson cloud as $m_\pi << m_\rho$. Nevertheless, the contribution of rho meson is non-negligible, and 
 both the contributions together add up in the whole $x$ region.
 The mesonic contribution dominates in the intermediate region of $x~ (0.2 < x < 0.6)$. For the medium nuclei like $^{4}He$, $^{12}C$, etc., mesonic contribution is small~\cite{SajjadAthar:2009cr}; while 
 it becomes pronounced in heavier nuclear targets like $^{40}$Ar, $^{56}$Fe, etc.~\cite{Haider:2012nf}. 
  In our earlier works we have found that in $F_{1A}(x,Q^2)$ and 
  $F_{2A}(x,Q^2)$ the mesonic contributions lead to an enhancement of the nuclear structure function, and it works in the 
 right direction to explain the experimental
 data~\cite{Marco:1995vb,SajjadAthar:2009cr,Haider:2015vea}. 
 
 
 Now to take into account the contribution from the virtual mesons, the neutrino self-energy is again evaluated using many body techniques~\cite{Marco:1995vb}. 
 For the mesonic effect we draw a diagram similar to Fig.\ref{wself_energy}, but here a nucleon propagator is replaced by a meson propagator. 
The meson propagator does not corresponds to the free mesons but it corresponds to the mesons which are arising due to the nuclear medium effects~\cite{FernandezdeCordoba:1991wf}. In the nuclear medium these mesons are arising  through particle-hole $(1p–1h)$, delta-hole $(1\Delta–1h)$,
 $1p1h-1\Delta1h$, $2p-2h$, etc. interactions as shown in Fig.\ref{n_self-1}. 
 
 The mesonic structure functions $F_{{i A,a}}(x_a,Q^2), ~~(i=1,2,5;a=\pi,\rho)$ are obtained as:
 \begin{eqnarray} 
\label{pion_f21}
F_{{i A,a}}(x_a,Q^2)  &=&  -6 \kappa \int \, d^3 r  \int \frac{d^4 p}{(2 \pi)^4} 
        \theta (p^0) ~\delta I m D_a (p) \;2m_a~\;f_{ia}(x_a),
\end{eqnarray}
 where
 \begin{eqnarray} 
\label{F1rho_wk}
f_{1a}(x_a) &=&  A m_a \left[\frac{F_{1a}(x_a)}{m_a}~+~\frac{{|{\bf p}|^2~-~(p^{z})^2}}{2(p^z~q^z~-~p^0~q^0)}
\frac{F_{2a}(x_a)}{m_a}\right],~~~~~
\end{eqnarray}
 \begin{eqnarray}
 \label{F2rho_wk}
 f_{2a}(x_a) &=&\left( \frac{F_{2a}(x_a)}{m_a^2 \nu}\right)\left[ \frac{Q^4}{q^0 {( q^z)}^2}\left(p^z+\frac{q^0(\gamma p^z-p^0) }{Q^2}{ q^z} \right)^2+\frac{q^0 Q^2 {(|{\bf p}|^2~-~(p^{z})^2)}}{2{( q^z)}^2}\right]
 \end{eqnarray}
 and
\begin{eqnarray}
\label{F5rho_wk}
f_{5N}(x_a)&=&A\;\;\frac{F_{5a}(x_a)}{m_a \nu}\times\left[q^0(\gamma p^z-p^0)+Q^2 \frac{p^z}{{ q^z}}. \right]
\end{eqnarray}
In Eqs.~\ref{pion_f21}, \ref{F1rho_wk}, \ref{F2rho_wk} and \ref{F5rho_wk}, $\kappa=1(2)$ for pion(rho meson), $\nu=\frac{q_0(\gamma p^z-p^0)}{m_a}$, $x_a=-\frac{Q^2}{2p \cdot q}$, $m_a$ is the mass of the meson($\pi$ or $\rho$). $D_a(p)$ is the meson($\pi$ or $\rho$) propagator in the nuclear medium and is written as 
 \begin{equation}\label{dpi}
D_a (p) = [ p_0^2 - {\bf {p}}\,^{2} - m_a^2 - \Pi_{a} (p_0, {\bf p}) ]^{- 1}\,,
\end{equation}
with
\begin{equation}\label{pionSelfenergy}
\Pi_a(p_0, {\bf p})=\frac{f^2}{m_\pi^2}\;\frac{C_\rho\;F^2_a(p){\bf {p}}\,^{2}\Pi^*}{1-{f^2\over m_\pi^2} V'_j\Pi^*}\,,
\end{equation}
where, $C_\rho=1(3.94)$ for pion(rho meson). $F_a(p)={(\Lambda_a^2-m_a^2) \over (\Lambda_a^2 - p^2)}$ is the $\pi NN$ or $\rho NN$ form 
factor, 
$\Lambda_a$=1 GeV (fixed by Aligarh-Valencia group~\cite{SajjadAthar:2009cr, Haider:2015vea}) and $f=1.01$.  $V_j'$ is the longitudinal(transverse)
part of the spin-isospin interaction for pion(rho meson), and $\Pi^*$ is the irreducible meson self-energy that contains the contribution of particle-hole and delta-hole excitations. 
 For the pions, we have used the PDFs parameterization given by Gluck et al.\cite{Gluck:1991ey} and for the $\rho$ mesons same PDFs as for the pions have been used as there
 is no available explicit parameterization for the $\rho-$meson PDFs in the literature. It is important
to mention that mesonic contribution does not play any role to $F_{3A}(x,Q^2)$. The reason is that $F_{3A}(x,Q^2)$ depends mainly on the valence quark distribution and not on the
sea quarks distribution. In the evaluation of $F_{4A}(x,Q^2)$, the mesonic contribution has not been incorporated because the mesonic PDFs for $F_{4A}(x,Q^2)$ are not 
available in the literature and for $F_{5A}(x,Q^2)$ mesonic effect is included by using the Albright-Jarlskog relation at the leading order as the parameterization for 
mesonic PDFs for $F_{2A}(x,Q^2)$ is available in the literature.

\subsubsection{Shadowing and Antishadowing effects}\label{shad}

The shadowing effect is taken into consideration following the works of Kulagin and Petti~\cite{Kulagin:2004ie,Kulagin:2007ju}, who have used the 
Glauber-Gribov multiple scattering theory.  In the case of $\nu_\mu/\bar\nu_\mu$ induced DIS processes, they have treated (anti)shadowing differently from the prescription applied 
in the case of electromagnetic structure functions~\cite{Kulagin:2004ie, Kulagin:2007ju}, due to the presence of the axial-vector  current  in  the neutrino interactions. 
The interference between the vector and the axial-vector currents introduces C-odd  terms  in  the neutrino   cross  sections,  which  are  described 
by structure function $F_3(x,Q^2)$. In their calculation of nuclear corrections, separate contributions to different structure functions according to 
their C-parity have been taken into account. This results in a different dependence of nuclear effects on the nuclear structure functions depending upon their C-parity specially 
in  the  nuclear  (anti)shadowing  region~\cite{Haider:2011qs}. We have adopted the same prescription for the inclusion of (anti)shadowing effect~\cite{Kulagin:2004ie,Kulagin:2007ju} in  
case of $\nu_\tau/\bar\nu_\tau-$nucleus scattering.

 \section{Results and Discussion}
  The present model describes the nuclear structure functions $F_{iA}(x_A,Q^2);~(i=1-5)$, 
 in terms of the nucleon structure functions $F_{iN}(x_N,Q^2);~(i=1-5)$, convoluted with the spectral function of the nucleon in the nucleus ($S_h$). The spectral function takes into account the effect 
 of Fermi motion, binding energy and nucleon correlations.  
 The results for the nucleon structure functions $F_{iN}(x_N,Q^2);~(i=1-5)$ at the leading order are obtained using nucleon PDFs of MMHT~\cite{Harland-Lang:2014zoa}. The structure functions are obtained in the three massless flavor ($n_f=3$) MSbar scheme as well as in the four 
 flavor ($n_f=4$) MSbar scheme, taking the charm quark mass into account ($m_c=1.3$ GeV~\cite{Kretzer:2003iu}). All the numerical evaluations have been performed for $Q^2>1$ GeV$^2$. 
 Then we evaluate the structure functions up to the next-to-leading order following the works of Kretzer et al.~\cite{Kretzer:2002fr}.
 The target mass correction has been included following Ref.~\cite{Kretzer:2003iu} and the dynamical higher twist (twist-4) correction has been taken into account  
 following the methods of Dasgupta et al.~\cite{Dasgupta:1996hh}. 
 In the numerical results, the HT effect is applied only on the three nucleon structure functions, i.e. $F_{iN}(x_N,Q^2);~(i=1-3)$ and is not explicitly applied on the massive charm quark. 
 Then the mesonic effects which include the contributions from the pion and the rho meson is taken into account 
 for $F_{iA}(x_A,Q^2);~(i=1,2,5)$ and (anti)shadowing effect is also included into the nuclear structure functions $F_{iA}(x_A,Q^2)$ for $i=$1, 2, 3 and 5.   
 For the mesonic cloud contribution we use the pionic PDFs parameterization of 
 Gluck et al.~\cite{Gluck:1991ey}.
 
 Let us first recapitulate the findings of our earlier works~\cite{Ansari:2020xne, Zaidi:2019asc, Ansari:2021cao}, for the free nucleon structure functions:
 \begin{itemize}
  \item The effect of  higher order perturbative evolution of parton densities at the next-to-leading order, is to increase the nucleon structure functions in the entire region of $x$.
 \item  The effect of target mass correction is to decrease the nucleon structure functions in the region of low $x$ up to $x \le 0.5$, after which it leads to an increase in the structure functions.
  \item The inclusion of higher twist corrections results in a small change ($<1-2\%$) in $2xF_{1N}(x,Q^2)$ and $F_{2N}(x,Q^2)$ evaluated at NLO, while 
  in $xF_{3N}(x,Q^2)$ there is a significant change in the entire range of $x$ in the region of low and moderate $Q^2$. Quantitatively, in $F_{3N}(x,Q^2)$ higher twist effect is found to be 
  $20\%(7\%)$ at $x=0.3$ and $21\%(11\%)$ at $x=0.8$ for $Q^2=2(5)$ GeV$^2$. 
  \item  The results of nucleon structure functions evaluated at NNLO with TMC effect are found to be close i.e. within  $<1\%$ to the results obtained at NLO with TMC and 
  higher twist corrections. 
  \item We find that the inclusion of tau lepton mass leads to a reduction in the differential scattering cross section which is predominantly due to the contribution from $F_{5N}(x,Q^2)$, in addition
  to the kinematical effect. 
  The contribution of $F_{4N}(x,Q^2)$ to the cross section is small.
 \end{itemize}

We now present the numerical results of the study performed in this work for $\nu_\tau(\bar\nu_\tau)-^{40}$Ar scattering with nuclear medium effects. 
In the numerical results `SF' corresponds to the case when the results, are obtained using only the spectral function, and the `Total' corresponds to 
the results of the full model, where the additional contributions from the meson clouds as well as the shadowing effects are taken into account.  
 The expression for the total nuclear structure functions in the present model is given by:
\begin{eqnarray}\label{f1f2_tot}
 F_{iA}(x,Q^2)=F_{iA,N}(x,Q^2)+F_{iA,\pi}(x,Q^2)+F_{iA,\rho}(x,Q^2)
 +F_{iA,shd}(x,Q^2),\;\;\;\;\;
\end{eqnarray}
where $i=1,2,5$. $F_{iA,N}(x,Q^2)$ are the nuclear structure functions which have contribution only from the spectral function and $F_{iA,\pi/\rho}(x,Q^2)$ is the
contribution from the mesons. 

$F_{iA,shd}(x,Q^2);~(i=1,2,5)$ have contribution from the 
shadowing effect given by~\cite{Kulagin:2004ie}
\begin{equation}\label{shad_sf}
 F_{iA,shd}(x,Q^2) = \delta R_i(x,Q^2) \times F_{i,N}(x,Q^2),
\end{equation}
where $\delta R_i(x,Q^2)$ are the shadowing correction factors. 

The full expression
for the parity violating weak nuclear structure function $F_{3A}(x,Q^2)$ is given by,
\begin{eqnarray}\label{f3_tot}
 F_{3A}(x,Q^2)= F_{3A,N}(x,Q^2) + F_{3A,shd}(x,Q^2).
\end{eqnarray}
From Eq.~\ref{f3_tot}, it may be noticed that $F_{3A}(x,Q^2)$ has no mesonic contribution as the contribution to this structure function comes mainly from the valence quarks ($u_v$ and $d_v$). 
For $F_{3A,shd}(x,Q^2)$ similar definition has been used~\cite{Haider:2011qs} as given
in Eq.(\ref{shad_sf}) following the works of Kulagin et al.~\cite{Kulagin:2004ie}. 

In view of $F_{4N}(x,Q^2)$ being very small as it vanishes in the leading order and contributes only due to higher order corrections we have evaluated $F_{4A}(x,Q^2)$ using only the spectral function, i.e., the contributions from the mesonic and shadowing effects have not been included, and therefore
\begin{equation}\label{f4_tot}
 F_{4A}(x,Q^2)=F_{4A,N}(x,Q^2).
\end{equation}
 \begin{figure}[ht]
 \includegraphics[height=10 cm, width=.9\textwidth]{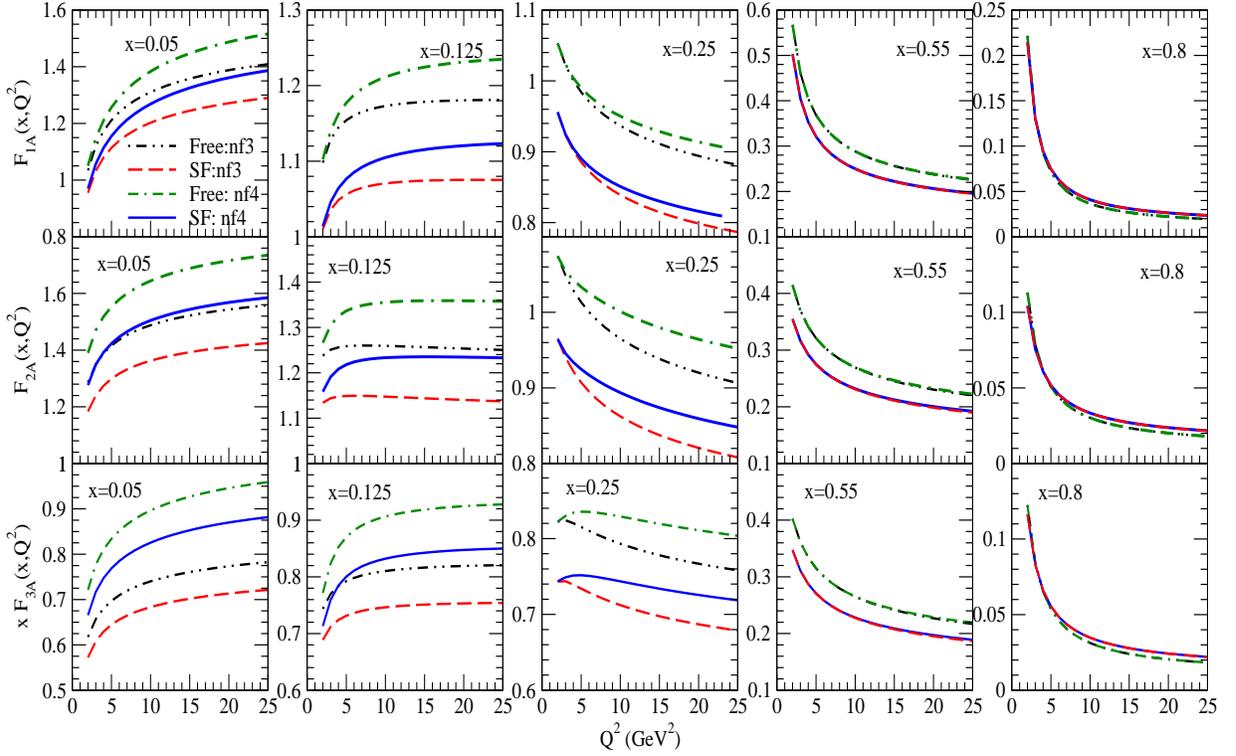}
\caption{Results for the nuclear structure functions $F_{iA}(x,Q^2)~;(i=1-3)$ obtained only with the spectral function vs $Q^2$ are shown at the different values of $x$. 
The results are obtained by treating $u,d,s$ quarks to be massless and $c$ quark to be massive. The numerical calculations are performed by 
incorporating TMC effect~\cite{Kretzer:2003iu} at NLO using MMHT nucleon PDFs parameterization~\cite{Harland-Lang:2014zoa}. 
$nf_3$ and $nf_4$ denote the evaluation of $F_{iA}(x,Q^2)$ in the three flavor ($u$, $d$, $s$) and four flavor ($u$, $d$, $s$ and $c$)
MSbar scheme, respectively. These results are also compared with the free nucleon case. For the present case no cut is applied on the center of mass energy $W$.}
\label{fig:sfaf}
 \end{figure}

The mesonic and the shadowing effects have been incorporated in $F_{5A}(x,Q^2)$ assuming the Albright-Jarlskog relation between $F_{5N}(x,Q^2)$ and $F_{2N}(x,Q^2)$ to 
be valid for the mesons also at the leading order, and we use the following expressions:
\begin{eqnarray}
F_{5A,\pi/\rho}(x,Q^2)&=&\frac{F_{2A,\pi/\rho}(x,Q^2)}{2x},
\label{f5_pirho}\\
 F_{5A,shd}(x,Q^2)&=&\frac{F_{2A,shd}(x,Q^2)}{2x}.
 \label{f5_shad}
\end{eqnarray}

Using Eqs.~\ref{f1f2_tot}-\ref{f5_shad} described above we have evaluated the nuclear structure functions $F_{iA}(x,Q^2)$ and using them obtained the differential 
scattering cross sections $\frac{1}{E_\nu}\frac{d^2\sigma}{dxdy}$ vs $y$ and $\frac{1}{E_\nu}\frac{d\sigma}{dy}$ vs $y$ by integrating over the Bjorken $x$.

In Fig.\ref{fig:sfaf}, we present the results for the nuclear structure functions viz. $2xF_{1A}(x,Q^2)$, $F_{2A}(x,Q^2)$ and $xF_{3A}(x,Q^2)$ (top to bottom) vs $Q^2$,
at the different values of $x$ lying in the range of $0.05\le x \le 0.8$
in the three-flavor as well as four-flavor MSbar scheme showing explicitly the effect of charm quark mass with $m_c=1.3$ GeV~\cite{Kretzer:2003iu}.
 The numerical calculations are performed at the next-to-leading order with the target mass corrections. 
The results of $F_{iA}(x,Q^2);~(i=1-3)$ obtained only with the spectral function have been compared with the corresponding 
results of free nucleon structure functions $F_{iN}(x,Q^2);~(i=1-3)$. It may be noticed that due to the presence of nuclear medium effects, the results of 
nuclear structure functions get suppressed from the results of the free nucleon case. For example, a suppression of $\approx 8\%(10\%)$ is found at $x=0.05(0.25)$ and at $Q^2=3$ GeV$^2$ in the results of nuclear
structure functions. 
From the figure, one may notice that effect of massive charm quark is important up to $x\le 0.2$ 
for the free nucleon as well as in the evaluation of nuclear structure functions like at $x=0.05$ we find an enhancement of $2\%$, $9\%$ and $18\%$ for 
$Q^2=3$ GeV$^2$, which becomes $8\%$, $11\%$ and $22\%$ for $Q^2=20$ GeV$^2$, respectively, in $F_{1N}(x,Q^2)$, $F_{2N}(x,Q^2)$ and $F_{3N}(x,Q^2)$. At $x=0.25$ massive charm 
effect is found to be $<1\%$ in all the three nucleon structure functions $F_{iN}(x,Q^2);(i=1-3)$ for $Q^2=3$ GeV$^2$ while for $Q^2=20$ GeV$^2$ it is found to be $\sim 2\%$ in 
$F_{1N}(x,Q^2)$, $5\%$ in $F_{2N}(x,Q^2)$ and $\sim 6\%$ in $F_{3N}(x,Q^2)$. It is important to notice that massive charm effect is more pronounced in $F_{3N}(x,Q^2)$
than in $F_{1N}(x,Q^2)$ and $F_{2N}(x,Q^2)$. We have observed that the contribution of massive charm quark to the nucleon as well as
the nuclear structure functions increases with the increase in $Q^2$ and decreases with the increase in $x$. 
Moreover, for the nuclear structure functions obtained only with the spectral function, the contribution of the charm quark is found to be approximately the same as we have observed
in the case of free nucleon structure functions. 
  \begin{figure}
 \includegraphics[height=6 cm, width=.95\textwidth]{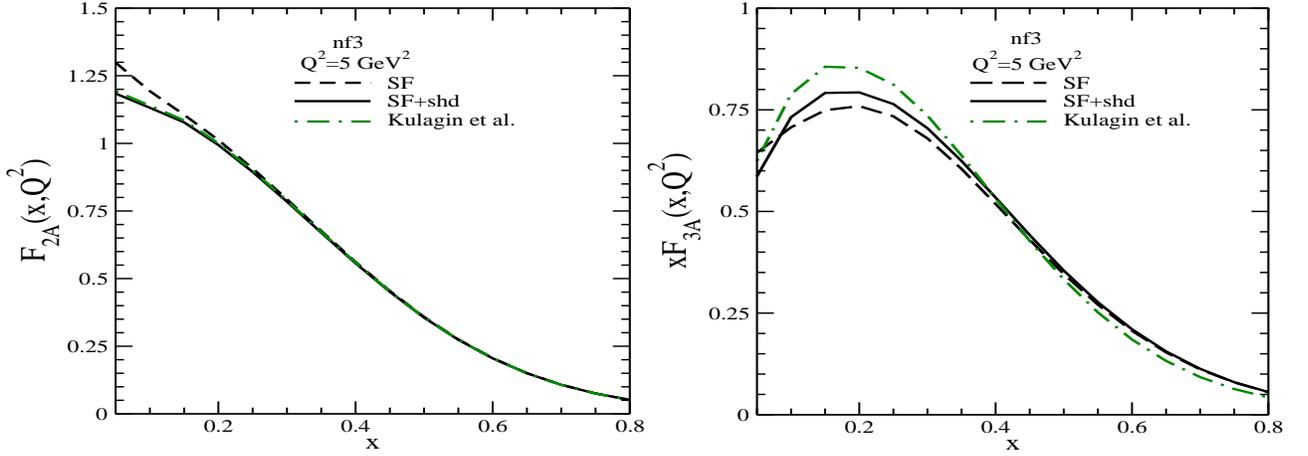}
 \caption{Results for the nuclear structure functions $F_{2A}(x,Q^2)$ and $xF_{3A}(x,Q^2)$ vs $x$ are shown at $Q^2=5$ GeV$^2$. 
$nf_3$ denotes the evaluation of $F_{iA}(x,Q^2);~(i=2,3)$ in the three flavor scheme by treating $u,d$ and $s$ quarks to be massless. The numerical 
calculations are performed by 
incorporating TMC effect~\cite{Kretzer:2003iu} at NLO using MMHT nucleon PDFs parameterization~\cite{Harland-Lang:2014zoa}. The results are 
compared with the results of Kulagin et al.~\cite{Kulagin:2004ie}.}
\label{fig:sfa0}
 \end{figure}
 
  In Fig.\ref{fig:sfa0}, we present the numerical results of $F_{2A}(x,Q^2)$ and $xF_{3A}(x,Q^2)$ vs $x$ at $Q^2=5$ GeV$^2$. 
  These results are obtained using the spectral function only (dashed line) and when the (anti)shadowing corrections are also included (solid line) in the three flavor MSbar scheme at NLO with TMC effect. 
 We have also compared the results with the numerical results of Kulagin et al.~\cite{Kulagin:2004ie} (dashed-dotted line). From the figure it may be observed that
 the results for $F_{2A}(x,Q^2)$ are in good agreement while our theoretical results for $xF_{3A}(x,Q^2)$ are $\sim 7\%$ lower from the results of Kulagin et al. at $x=0.2$, however, for $x>0.3$
we find them to be in reasonable agreement.

 \begin{figure}[h]
 \includegraphics[height=12 cm, width=.95\textwidth]{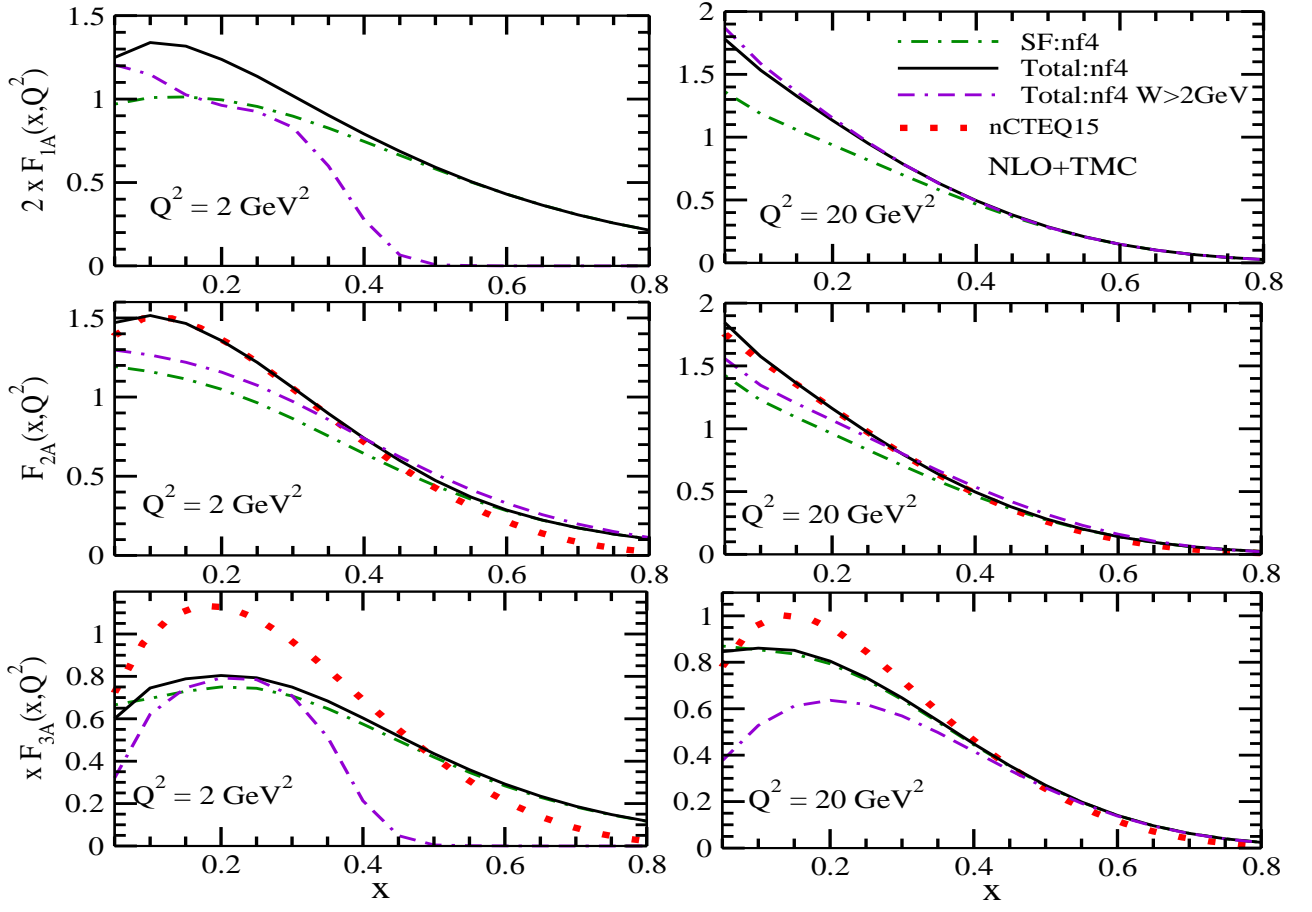}
 \caption{Results for the nuclear structure functions $F_{iA}(x,Q^2)~;(i=1-3)$ vs $x$ are shown for the different values of $Q^2$. 
$nf_4$ denotes the evaluation of $F_{iA}(x,Q^2)$ in the four flavor scheme ($u$, $d$, $s$ and $c$) by treating $u,d,s$ quarks to be massless and $c$ quark to be massive. The numerical 
calculations are performed by 
incorporating TMC effect~\cite{Kretzer:2003iu} at NLO using MMHT nucleon PDFs parameterization~\cite{Harland-Lang:2014zoa}. The results are also presented for 
the case when a cut on the CM energy $W>2$ GeV is applied. 
`SF' corresponds to the results only with the spectral 
function, and `Total' corresponds to the results of full model, where the additional contribution from the mesonic clouds and the shadowing effects are taken into account. The results of 
$F_{2A}(x,Q^2)$ and $xF_{3A}(x,Q^2)$ are compared with the phenomenological results obtained using the nCTEQ15 nuclear PDFs parameterization~\cite{Kovarik:2015cma}.}
\label{fig:sfa}
 \end{figure}

 In Fig.\ref{fig:sfa}, the results for the nuclear structure functions $2xF_{1A}(x,Q^2)$ (top), $F_{2A}(x,Q^2)$ (middle) and $xF_{3A}(x,Q^2)$ (bottom) vs $x$  are shown at NLO with TMC effect. 
 The results are obtained at $Q^2=2$ GeV$^2$(left) and $Q^2=20$ GeV$^2$(right), without and with a CM energy cut of 2 GeV. The results presented here are relevant to understand the nuclear medium modifications, dependence on the kinematic variables such as $x$, $Q^2$ and $W$. 
 In general, the structure functions in the nuclear environment get reduced due to the effects of the spectral 
 function (in the entire range of $x$) and the shadowing correction (in the low $x(\le 0.1)$ region), while they get enhanced due to the mesonic contribution (up to mid $x\le 0.6$). 
 One may notice from the figure that the results obtained with the full theoretical model (`Total') which has 
 contribution from the spectral function, mesonic cloud and (anti)shadowing effects (Eq.\ref{f1f2_tot}) get enhanced as compared to the results obtained only with the spectral function(`SF') in 
 the case of $F_{1A}(x,Q^2)$ and $F_{2A}(x,Q^2)$. Quantitatively, the enhancement in
the results of nuclear structure functions with the full theoretical model from the results obtained only with the spectral function in $F_{1A}(x,Q^2)$ is about $\approx 30\% (33\%)$ at $x=0.05$,
$24\% (21\%)$ at $x=0.2$, $\approx 2\% (2 \%)$ at $x=0.5$ for $Q^2=2 (20)$ GeV$^2$ while in $F_{2A}(x,Q^2)$ this enhancement becomes 
 $23\% (25\%)$ at $x=0.05$, $29\% (21\%)$ at $x=0.2$, $\sim 8\% (\approx 3\%)$ at $x=0.5$ for $Q^2=2 (20)$ GeV$^2$.
However, in the case of $F_{3A}(x,Q^2)$ structure function, where there is no mesonic effect, we have observed that due to the shadowing corrections at very low $x$, for example, at $x=0.05$
 there is a further reduction from the results
 obtained using only the spectral function, which is about $11\%$ and $3\%$ at $Q^2=2$ GeV$^2$ and 20 GeV$^2$, respectively. Whereas
  at $x=0.1$ we observe an enhancement arising due to antishadowing correction which is about $7\%$ for $Q^2=2$ GeV$^2$ and it decreases to $\approx 1\%$ for $Q^2=20$ GeV$^2$.

 The present theoretical model was first applied to study the nuclear medium effects in the electromagnetic nuclear structure functions, i.e. $F_{1A}^{EM}(x,Q^2)$ and $F_{2A}^{EM}(x,Q^2)$ using
 different nuclear targets such as
beryllium, carbon, aluminium, calcium, iron, copper, tin, gold and lead~\cite{SajjadAthar:2009cr, Haider:2015vea, Haider:2016zrk, Zaidi:2019mfd}. These theoretical results were compared 
with the available data from EMC~\cite{EuropeanMuon:1986xsr}, SLAC~\cite{Gomez:1993ri}, NMC~\cite{NewMuon:1996yuf, NewMuon:1996fwh} and JLab~\cite{Seely:2009gt, Mamyan:2012th} experiments and were found to be in reasonable agreement. Moreover, in Ref.~\cite{Zaidi:2019mfd}, a comparative study of our theoretical results
with the phenomenological parameterizations of Whitlow et al.~\cite{Whitlow:1990gk, Whitlow:1991uw} and nCTEQ15 nuclear PDFs~\cite{Kovarik:2015cma} were made. In Refs.~\cite{Haider:2011qs, Haider:2012nf, Haider:2016zrk, Zaidi:2019asc}, 
this model was applied to understand the nuclear medium effects in $\nu_\mu(\bar\nu_\mu)-A$ DIS process for carbon, hydrocarbon, argon,
iron and lead nuclear targets which are presently being used in most of the (anti)neutrino oscillation experiments or those being used by the MINERvA collaboration in order to 
understand the hadron dynamics in the nuclear medium. The results of weak nuclear structure functions and the differential scattering cross sections were compared with 
the available experimental data of CDHSW~\cite{Berge:1989hr}, NuTeV~\cite{Tzanov:2005kr}, CCFR~\cite{Oltman:1992pq}, CHORUS~\cite{Onengut:2005kv} and MINERvA~\cite{Mousseau:2016snl} collaborations as
well as with the phenomenological parameterizations of nCTEQnu nuclear PDFs~\cite{morfin_private}, Hirai et al.~\cite{Hirai:2007sx}, Eskola et al.~\cite{Eskola:1998df}, Cloet et al.~\cite{Cloet:2006bq},
Bodek et al.~\cite{Bodek:2002vp, Bodek:2010km} and GENIE Monte Carlo generator~\cite{Andreopoulos:2009rq}. In the present work, the numerical results of $F_{2A}(x,Q^2)$ and $xF_{3A}(x,Q^2)$ in argon 
have been compared with the results obtained using the nCTEQ15 nuclear 
PDFs parameterization~\cite{Kovarik:2015cma} as shown in Fig.~\ref{fig:sfa}. It may be noticed from the figure that the results of $F_{2A}(x,Q^2)$ are consistent with the 
phenomenological results of nCTEQ15~\cite{Kovarik:2015cma} while the results of $xF_{3A}(x,Q^2)$ are different in the intermediate region of $x$ ($\le 0.4$), however, this difference 
decreases with the increase in $x$ and $Q^2$.

  We have also observed that the inclusion of $W$ cut suppresses the nuclear structure functions as shown in Fig.~\ref{fig:sfa}. The effect of kinematical cut on $W$ is summarized below:
  \begin{enumerate}[(i)]
    \item Due to the effect of CM energy cut of 2 GeV, i.e. $W>2$ GeV, the suppression in the results of $F_{1A}(x,Q^2)$ is found to be $14\%(3\%)$ at $x=0.1$ and $18\%(<1\%)$ at $x=0.3$, while
  in $F_{2A}(x,Q^2)$ it is found to be about $13\%(<1\%)$ at $x=0.1$ and $\sim 25\%(<1\%)$ at $x=0.3$ for $Q^2=2(20)$ GeV$^2$.
  In $F_{1A}(x,Q^2)$ and $F_{2A}(x,Q^2)$ the CM energy cut is important only in the low $Q^2$ region, and
  this effect becomes almost negligible for $Q^2>20$ GeV$^2$. It may be noticed that this suppression is $x$ dependent (large suppression at higher values of $x$).
  \item The nature of suppression in $F_{3A}(x,Q^2)$ is different from $F_{1A}(x,Q^2)$ and $F_{2A}(x,Q^2)$ and is significant even at high $Q^2$. Furthermore, it may be observed from Fig.~\ref{fig:sfa} 
  that the $x$ dependence of $F_{3A}(x,Q^2)$ is also different from $F_{1A}(x,Q^2)$ and $F_{2A}(x,Q^2)$, and the effect of $W$ cut is prominent at low $x$ even for $Q^2\sim 20$ GeV$^2$. 
  For example, at $x=0.1$ the suppression in the results of $F_{3A}(x,Q^2)$ with CM energy cut of $W>2$ GeV as compared to the results obtained without having any constrain
  on the CM energy is about $16\%$ for $Q^2=2$ GeV$^2$ and $38\%$ for $Q^2=20$ GeV$^2$ and at $x=0.3$ it becomes $5\%$ for $Q^2=2$ GeV$^2$ and $12\%$ for $Q^2=20$ GeV$^2$.
 
   \end{enumerate}
  \begin{figure}
 \includegraphics[height=10 cm, width=.95\textwidth]{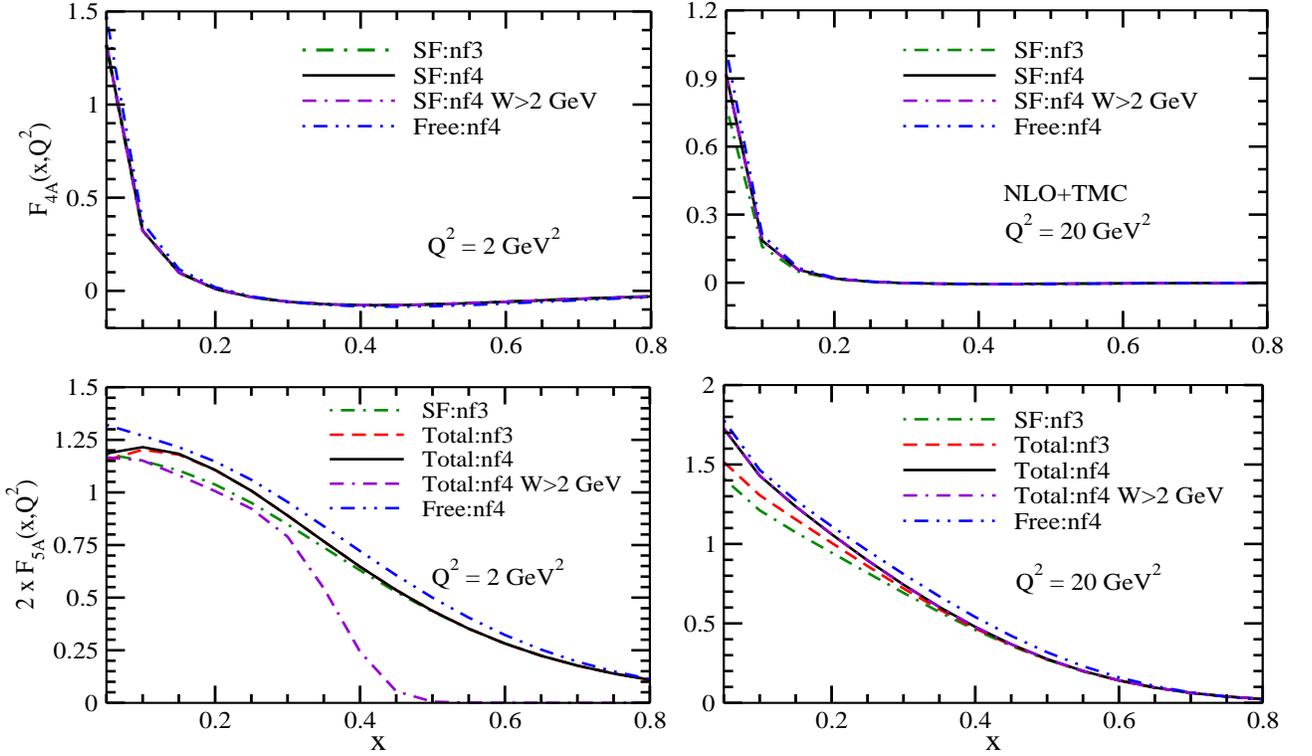}
\caption{Results for nuclear structure functions $F_{iA}(x,Q^2)~;(i=4,5)$ vs $x$ are shown at the different values of $Q^2$. 
The results are obtained by treating $u,d,s$ quarks to be massless and $c$ quark to be massive. The numerical calculations are performed by 
incorporating TMC effect~\cite{Kretzer:2003iu} at NLO using MMHT nucleon PDFs parameterization~\cite{Harland-Lang:2014zoa}. 
$nf_3$ and $nf_4$ denote the evaluation of $F_{iA}(x,Q^2)$ in the three flavor ($u$, $d$, $s$) and four flavor ($u$, $d$, $s$ and $c$)
MSbar scheme, respectively. `SF' corresponds to the results only with the spectral 
function, and `Total' corresponds to the results of full model, where the additional contribution from the mesonic clouds and the shadowing effects are taken into account. }
\label{fig:sfa1}
 \end{figure}
 
   \begin{figure}
 \includegraphics[height=8 cm, width=.99\textwidth]{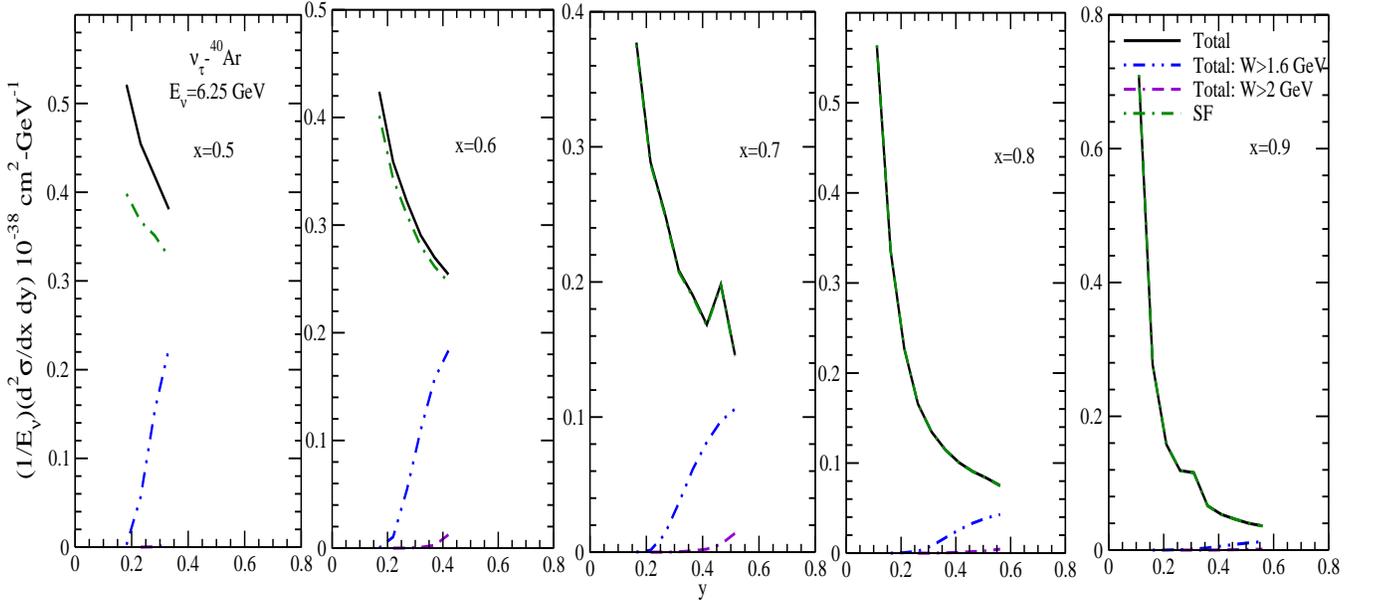}
\caption{Results for the double differential scattering cross section, $\frac{1}{E_\nu}\frac{d^2\sigma_A}{dx dy}$ vs $y$, are shown at the different values of
$x$ for $E_\nu=6.25$ GeV in $\nu_\tau-^{40}Ar$. `SF' corresponds to the results only with the spectral 
function, and `Total' corresponds to the results of full model, where the additional contribution from the mesonic clouds and the shadowing
effects are taken into account. The curves shown above have been obtained by incorporating TMC~\cite{Kretzer:2003iu} 
and HT~\cite{Dasgupta:1996hh} effects at NLO in the four flavor scheme (massless quarks viz. $u$, $d$, $s$ and massive $c$ quark). 
For the numerical calculations MMHT nucleon PDFs parameterization~\cite{Harland-Lang:2014zoa} has been used.
Dashed-dotted and solid lines represent the result of cross section with only the 
spectral function and with the full model, respectively. Dash double-dotted and double-dash dotted lines, respectively show the results for full model with $W>1.6$ GeV and $W>2.0$ GeV. }
\label{fig_d2sig5gev}
 \end{figure}
  In Fig.\ref{fig:sfa1}, the results for the nuclear structure functions $F_{4A}(x,Q^2)$ and $F_{5A}(x,Q^2)$ vs $x$, are shown, considering all the cases discussed above 
  for Fig.\ref{fig:sfa}. These are the two additional structure functions which contribute to the charged current $\nu_\tau/\bar\nu_\tau-$nucleus scattering cross sections in the 
  case of $m_\tau \ne 0$ and their contributions are negligible in $\nu_e/\nu_\mu$ induced charged current DIS. Here we find that $F_{4A}(x,Q^2)$ has a finite contribution in the region of 
  low $x(\le 0.2)$ and at low $Q^2~(Q^2\sim 2-5)$ GeV$^2$, while at higher values of $x$ 
   its contribution becomes almost negligible. 
   For example, when the evaluation of PDFs is performed at the next-to-leading order, we have observed that the value of $F_{4N}(x,Q^2)$ is finite and considerably large at 
   very low $x$ as compared to the leading order case, where $F_{4N}(x,Q^2)=0$.
  Moreover, we find that in the case of bound nucleon the results of
  nuclear structure functions $F_{4A}(x,Q^2)$ gets suppressed by about $10\%-12\%$ due to nuclear medium effects as compared to the results of 
  $F_{4N}(x,Q^2)$, in the region of $x\le0.2$. With the increase in $Q^2$ ($20$ GeV$^2$ vs $2$ GeV$^2$),  $F_{4A}(x,Q^2)$ contributes only at very low $x$. The effects
  of $W$ cut and charm mass are found to be small in $F_{4A}(x,Q^2)$. For $F_{5A}(x,Q^2)$, we have noticed that the $x$ and $Q^2$ dependence is qualitatively similar as observed in
  the case of $F_{2A}(x,Q^2)$. The effect of nuclear corrections obtained only with the spectral function has been found to be qualitatively similar in $F_{5A}(x,Q^2)$ 
  and $F_{2A}(x,Q^2)$, and the mesonic cloud contributions in $F_{5A}(x,Q^2)$, which is incorporated using the Albright-Jarlskog relation at the leading order
  give rise to an enhancement in the nuclear structure function. For example, at $Q^2=2(20)$ GeV$^2$ the mesonic cloud contribution is found
  to be $\sim 7\%(6\%)$ at $x=0.2$ and $<1\%(<1\%)$ at $x=0.5$. By performing a comparative study of $F_{2A}(x,Q^2)$ and $F_{5A}(x,Q^2)$, we find that Albright-Jarlskog
  relation gets violated due to the presence of nuclear medium effects (not shown here explicitly), especially in the region of low and intermediate $x(\le 0.6)$ and 
  with the increase in $x$ and $Q^2$ the difference between $F_{2A}(x,Q^2)$ and $2 x F_{5A}(x,Q^2)$ becomes almost negligible. Other effects like the inclusion of 
  massive charm quark or kinematical constrain (CM energy cut) have been found to be qualitatively similar to what has been observed in the case of $F_{2A}(x,Q^2)$.

 Using the results of the nuclear structure functions ($F_{iA}(x,Q^2);~(i=1-5)$), we evaluate the differential scattering cross section (Eq.~\ref{xsecsf}). All the numerical results are 
 obtained for $Q^2\ge1.0$ GeV$^2$ at NLO with HT and TMC effects in the energy range for $6.25 \le E_\nu \le 20$ GeV, which is the relevant energy region of 
 the present and future (anti)neutrino experiments. The effects of CM energy cut of $W>1.6$ GeV and $W>2$ GeV on the scattering cross sections have been also studied.
 
 In Fig.\ref{fig_d2sig5gev}, the results for the double differential scattering cross section $\frac{1}{E_\nu}\frac{d^2\sigma}{dx dy}$ vs $y$ is shown for the 
 different values of $x$ at $E_\nu=6.25$ GeV.
 We find that the contribution to the cross section comes from the intermediate and high region of $x$, and for $y \le 0.6$. In the presently considered kinematical region
 of $0.5 \le x \le 0.9$ and $y\le0.6$, the mesonic cloud contribution to the differential scattering cross section is significant in the region of low inelasticity $y$, however, it becomes small
 with the increase in $y$. For example, at $x=0.5$ there is an enhancement of $23\%$ for $y=0.2$ and $16\%$ for $y=0.3$
 in the full model as compared to the results obtained using only the spectral function. The kinematic region of $0.3\le Q^2\le6$ GeV$^2$ is sensitive to the nonperturbative QCD 
 corrections of higher twist effect, the inclusion of which along with the TMC effect leads to an enhancement 
 of about $21\%$ and $5\%$ at $x=0.6$ for $y=0.2$ and $y=0.4$, respectively
  as compared to the results obtained only with the TMC effect(not shown here explicitly). 
      \begin{figure}[h]
 \includegraphics[height=10 cm, width=.95\textwidth]{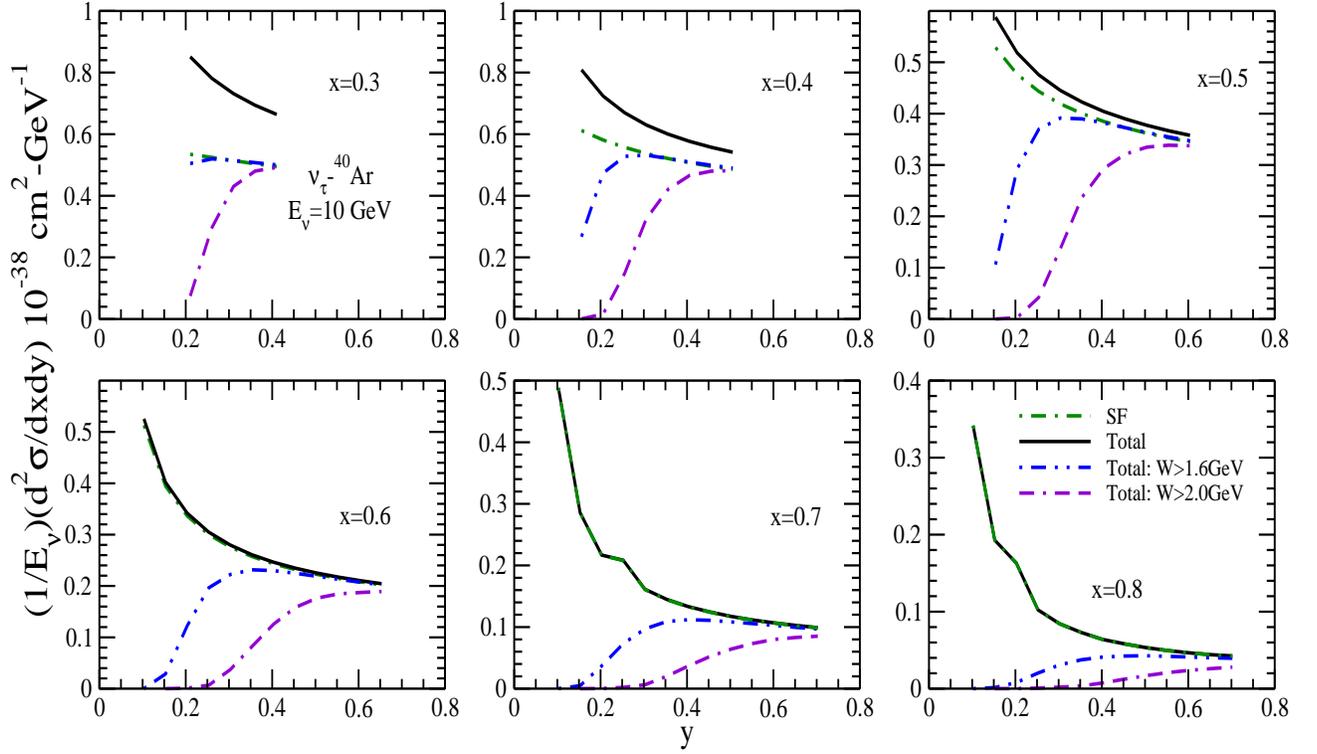}
\caption{Results for the double differential scattering cross section, $\frac{1}{E_\nu}\frac{d^2\sigma_A}{dx dy}$ vs $y$, are shown at the different values of
$x$ for $E_\nu=10$ GeV in $\nu_\tau-^{40}Ar$. The lines and symbols have the same meaning as in Fig.~\ref{fig_d2sig5gev}.}
\label{fig_d2sig10gev}
 \end{figure}
 We have found that the enhancement in the cross sections due to the HT effect becomes more pronounced with the increase in $x$, e.g., at $x=0.8$,  
 it is found to be $88\%$ for $y=0.2$ and $28\%$ for $y=0.4$.
 Furthermore, incorporation of the CM energy cut in the numerical calculations further 
 reduces the DIS cross section 
 like at $x=0.6$ and $y=0.4$ (corresponding to $Q^2=\sim 3$ GeV$^2$) the results of $\frac{1}{E_\nu}\frac{d^2\sigma}{dx dy}$ 
 obtained without any cut on the CM energy are $28\%$ and $95\%$ higher in magnitude as compared to the results with cuts of $W>1.6$ GeV and $W>2$ GeV, respectively. It is 
 important to point out that
 the application of $W\ge 2$ GeV cut (shown with double dash dotted lines), leads to a very small (almost negligible) contribution of DIS cross section in the considered kinematic region. 

\begin{figure}[h]
 \includegraphics[height=10 cm, width=.95\textwidth]{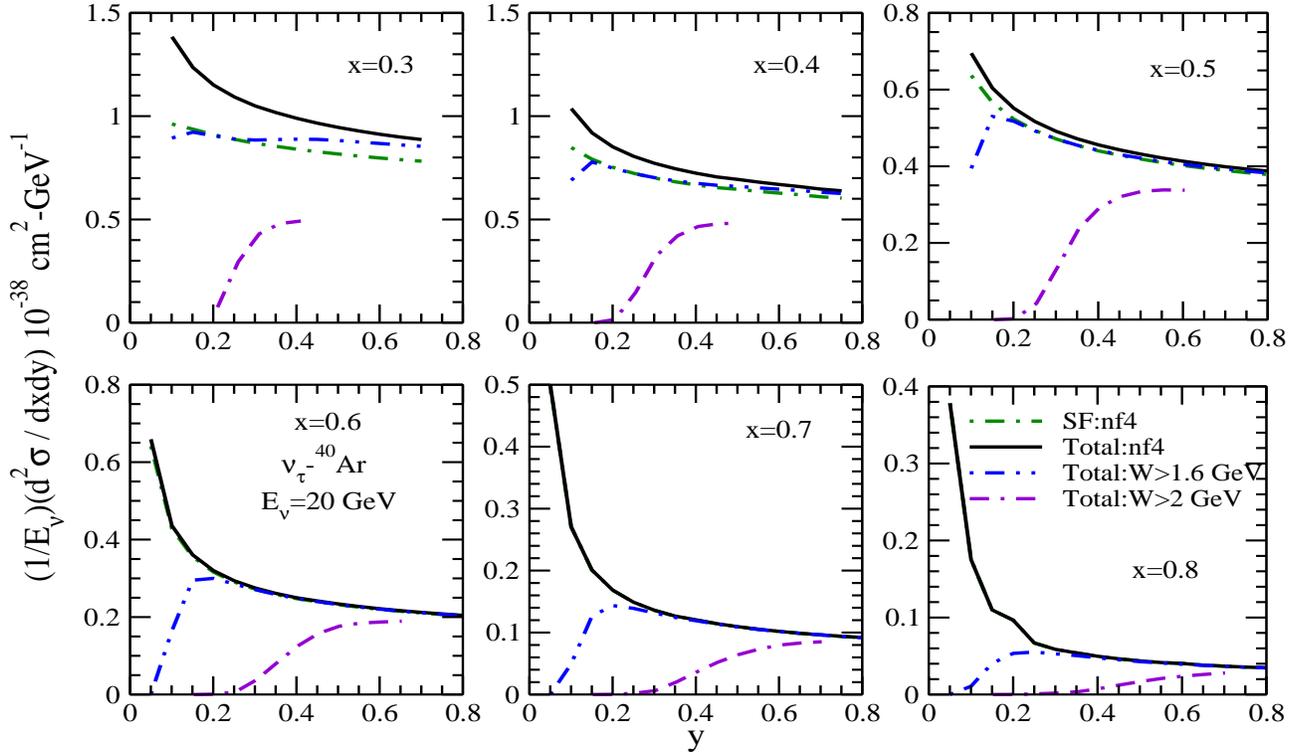}
\caption{Results for the double differential scattering cross section, $\frac{1}{E_\nu}\frac{d^2\sigma_A}{dx dy}$ vs $y$, are shown at the different values of
$x$ for $E_\nu=20$ GeV in $\nu_\tau-^{40}Ar$. The lines and symbols have the same meaning as in Fig.~\ref{fig_d2sig5gev}.}
\label{fig_d2sig20gev}
 \end{figure}

 To study the energy dependence of the cross section, we have  calculated the differential scattering cross sections at $E_\nu=10$ GeV as well as at $E_\nu=20$ GeV
 and the corresponding results are presented in Figs.~\ref{fig_d2sig10gev} and \ref{fig_d2sig20gev}, respectively. One may notice from the figures
 that with the increase in energy the differential cross section gets enhanced. For example, we find an enhancement of about $24\%(30\%)$ at $x=0.3$ and $18\%(17\%)$ 
 at $x=0.6$ for $y=0.2(0.4)$ in the results of cross section obtained at $E_\nu=20$ GeV as compared to the results obtain at $E_\nu$=10GeV. It is important to point out
 that up to $E_\nu=$20 GeV, the contribution to the cross section from the charm quark is negligible (not shown here explicitly).
 Moreover, we have found that the effect of twist-4 contribution (HT effect) decreases with the increase in energy, quantitatively as we move from $E_\nu=6.25$ GeV to 
 $E_\nu=10$ GeV, a reduction of about $7\%(1\%)$ for $y=0.2(0.4)$ at $x=0.6$ is found which becomes $43\%(14\%)$ at $x=0.8$. The impact of HT corrections is  
 further reduced for $E_\nu=20$ GeV. 
The inclusion of the CM energy cuts ($W>1.6$ GeV and $W>2.0$ GeV) significantly reduces the cross section, however, this reduction becomes small with the increase in energy in the wide kinematic
region of $x$ and $y$. For $E_\nu=10$ GeV, we find a reduction of about $25\%$($40\%$) with $W>1.6$ GeV cut 
(shown by dashed double-dotted line) which becomes $26\%$($90\%$) with $W>2$ GeV cut (shown by double dashed-dotted line) for $y=0.2(0.4)$ at $x=0.3$.

    \begin{figure}[h]
 \includegraphics[height=10 cm, width=.95\textwidth]{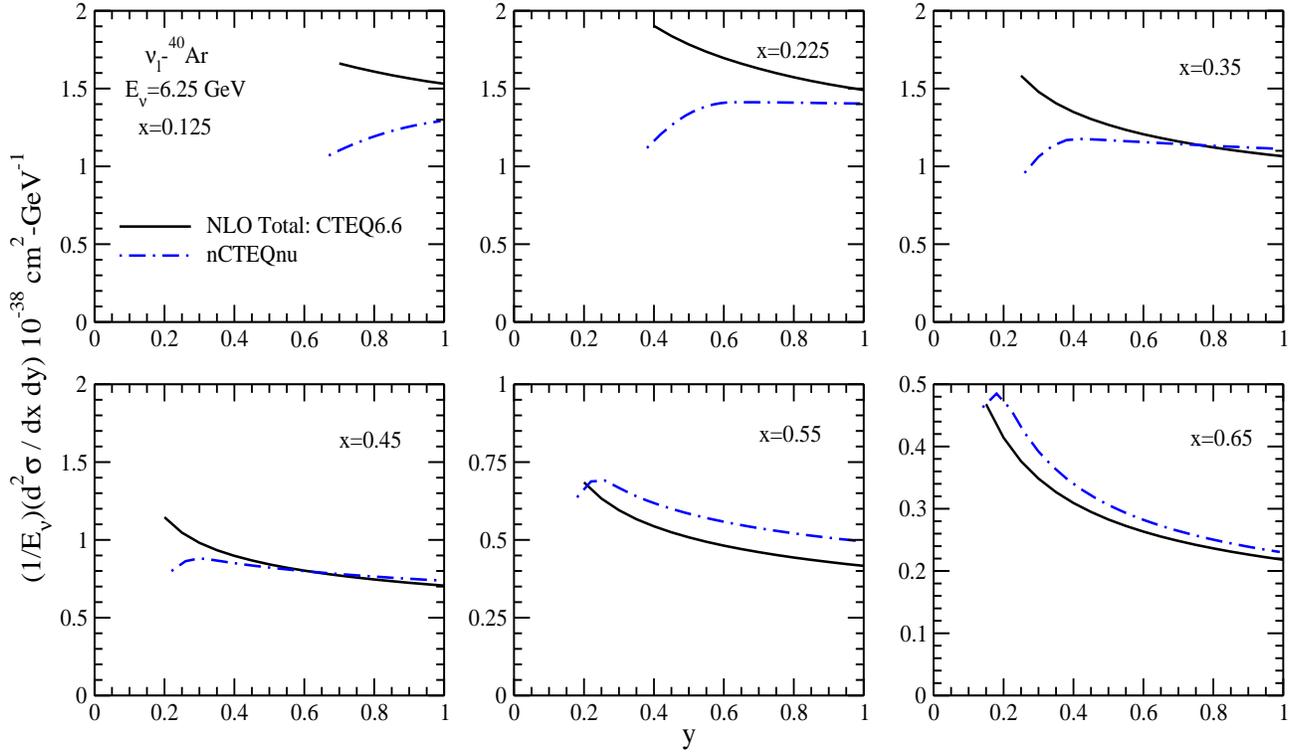}
\caption{Predictions for the differential scattering cross section vs $y$, at different values of $x$ for $E_\nu=$6.25 GeV in $\nu_\mu-^{40}Ar$.  The results are 
obtained with a constrain on $Q^2\ge1.0$ GeV$^2$ by using CTEQ6.6 nucleon PDFs~\cite{Nadolsky:2008zw} at NLO in the MS-bar scheme (solid line). Dash-dotted line represents the nCTEQnu~\cite{morfin_private} nuclear PDFs based prediction.}
\label{fig_d2sig6gev}
 \end{figure}
 
\begin{figure}[h]
 \includegraphics[height=9 cm, width=.85\textwidth]{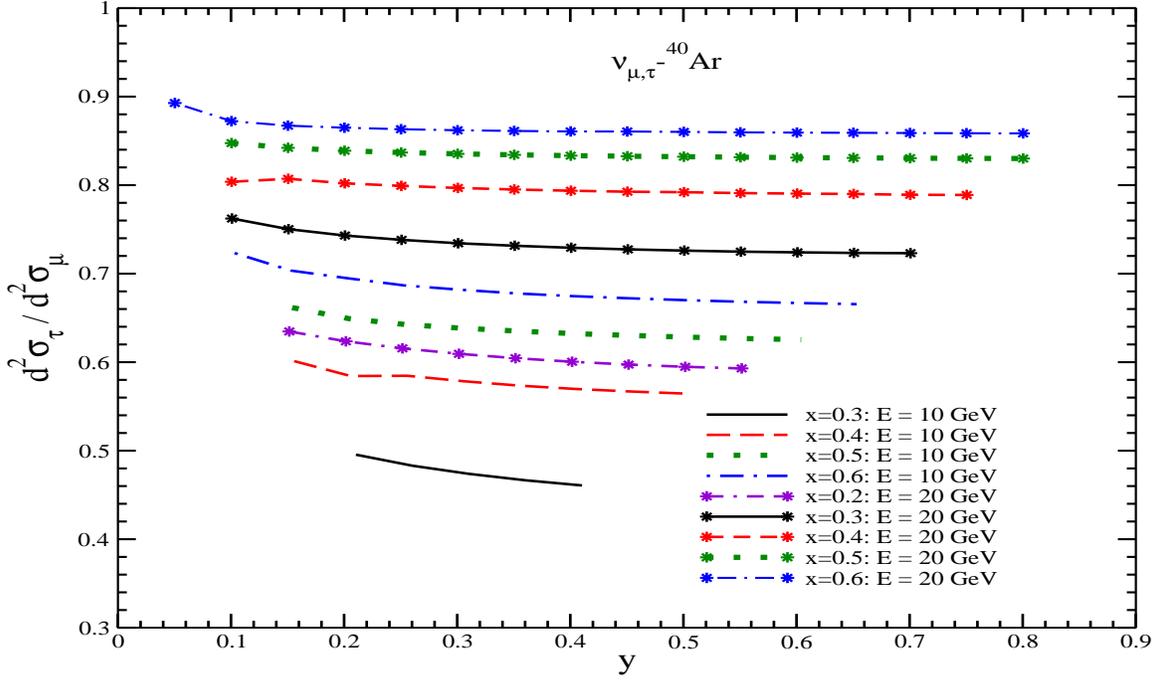}
\caption{Ratio of the differential scattering cross section $\frac{d^2 \sigma_{\nu_\tau}}{d^2 \sigma_{\nu_\mu}}$ vs $y$ without any cut on CM 
energy is shown~\cite{Kretzer:2002fr, Jeong:2010nt} at $E_\nu=10$ GeV and $E_\nu=20$ GeV. These results are obtained at NLO by using MMHT nucleon PDFs 
parameterization~\cite{Harland-Lang:2014zoa}. The effects of TMC~\cite{Kretzer:2003iu} and HT~\cite{Dasgupta:1996hh} are also included.}
\label{fig:ratio_d2sig}
 \end{figure}

To understand the impact of charged lepton mass on the scattering cross section in order to interpret the experimental data, we have also performed 
the numerical calculations for $\nu_\mu-^{40}Ar$ deep inelastic scattering cross section,
 where the contributions from $F_{4A}(x,Q^2)$ and $F_{5A}(x,Q^2)$ is negligible. Moreover, a comparison of the differential cross section obtained for the $\nu_\mu-^{40}Ar$ vs $\nu_\tau-^{40}Ar$ 
scattering processes has also been made to quantify the effect of lepton mass. These results are presented in Figs.~\ref{fig_d2sig6gev} and \ref{fig:ratio_d2sig}, respectively.
 
   The results of $\frac{1}{E_\nu}\frac{d^2\sigma_A}{dx dy}$ vs $y$ are presented in Fig.\ref{fig_d2sig6gev} at the different values of
 $x$ for $E_{\nu_\mu}$=6.25 GeV.
 For the sake of completeness, we have made a comparison of these theoretical results for 
 $\nu_\mu-^{40}Ar$ scattering cross section with the results obtained using the phenomenological nuclear PDFs prescribed by nCTEQnu collaboration~\cite{morfin_private} to
 obtain the cross sections. This comparative study gives an overview of existing uncertainties in the prediction of the cross sections.
 We observe that the present theoretical results with the TMC effect at NLO in 
 the four flavor MSbar scheme (solid line)~\cite{Nadolsky:2008zw} show significant
 deviation from the results obtained using the nCTEQnu nuclear PDFs parameterization~\cite{morfin_private} specially in the region of high $y$ and low $x$. However, in the intermediate range of $x$, i.e.,
 $0.35\le x\le 0.45$ (presented here) both the approaches are in reasonable agreement. It implies that in the region of few GeV ($<10$ GeV) more theoretical 
 as well as phenomenological efforts are required
 in order to develop a better understanding of neutrino interactions.
 
 In order to see the effect of finite lepton mass on the cross section, in Fig.~\ref{fig:ratio_d2sig}, we present the results for the ratio of the differential cross sections
 $\frac{\left(\frac{d^2 \sigma_{\nu}}{dxdy}\right)_{\nu_\tau-^{40}Ar}}{\left(\frac{d^2 \sigma_{\nu}}{dxdy}\right)_{\nu_\mu-^{40}Ar}}$ vs $y$, obtained 
 using the four flavor MSbar scheme at NLO with TMC and HT effects at the  different values of $x$, for  $E_\nu=10$ GeV and $E_\nu =20$ GeV.
%
 It may be noticed from the figure that the effect of the lepton mass decreases with the increase in energy and the ratio 
 $r\Big[=\frac{\left(\frac{d^2 \sigma_{\nu}}{dxdy}\right)_{\nu_\tau-^{40}Ar}}{\left(\frac{d^2 \sigma_{\nu}}{dxdy}\right)_{\nu_\mu- ^{40}Ar}}\Big]$ approaches unity at high energies.
 For example, at $x=0.3$ and $y=0.2(0.4)$, $r$ increases by  $24\%( 27\%)$  when we increase the projectile beam energy from 10 GeV to 20 GeV while it becomes $17\%(18 \%)$ at $x=0.6$.
  Furthermore, we observe that the ratio $r$ shows $x$ as well as $y$ dependence as the effect of lepton mass increases with the increase in $y$ and decrease in $x$.
 For example, at $y=0.4$ and $E_\nu=10(20)$ GeV the ratio increases by $21\%(18\%)$ when $x$ is varied from 0.3 to 0.6. 
These results would be relevant for the upcoming DUNE experiment, where $\nu_\mu \to \nu_\tau$ oscillation channel is planned to be studied.

\begin{figure}[h]
 \includegraphics[height=8 cm, width=.99\textwidth]{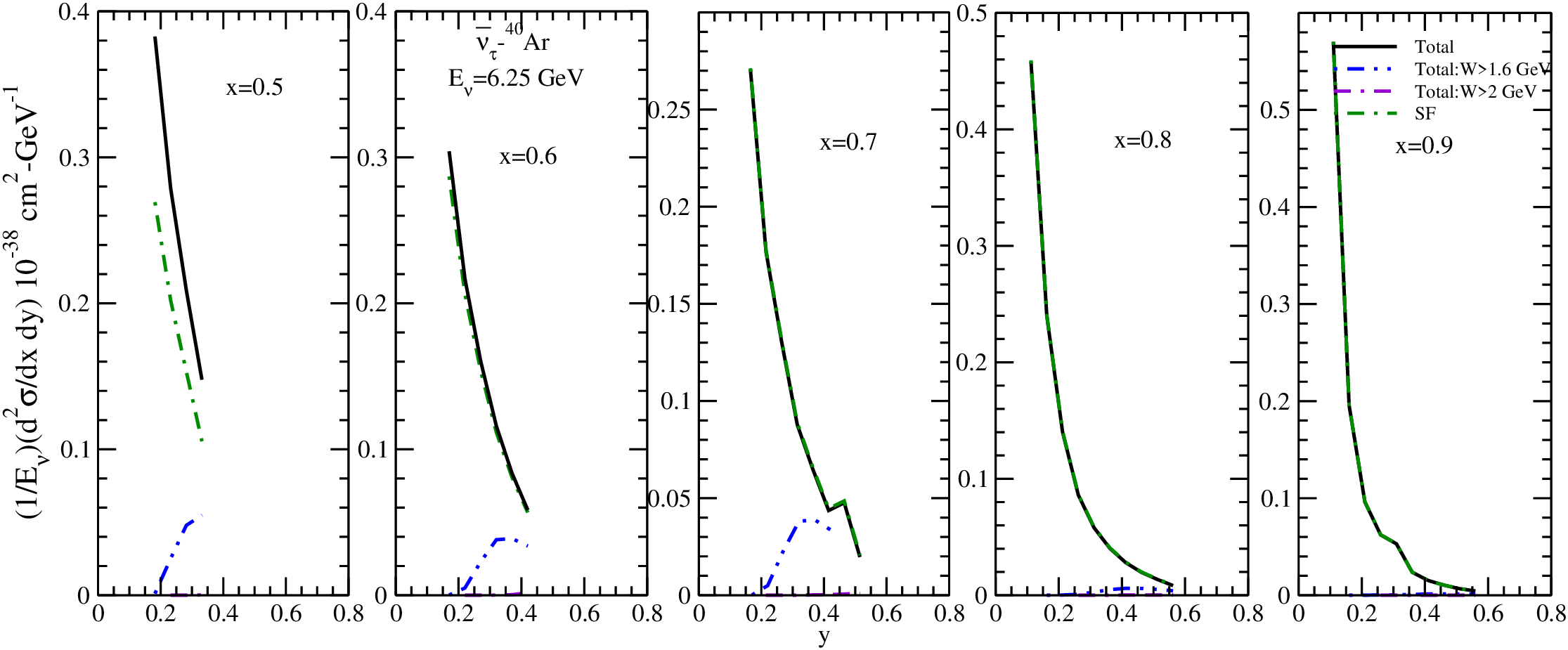}
\caption{Results for the double differential scattering cross section, $\frac{1}{E_\nu}\frac{d^2\sigma_A}{dx dy}$ vs $y$, are shown at the different values of
$x$ for $E_\nu=6.25$ GeV in $\bar\nu_\tau-^{40}Ar$. The lines and symbols have the same meaning as in Fig.~\ref{fig_d2sig5gev}.}
\label{fig_d2sig5geva}
 \end{figure}
 
\begin{figure}
 \includegraphics[height=10 cm, width=.95\textwidth]{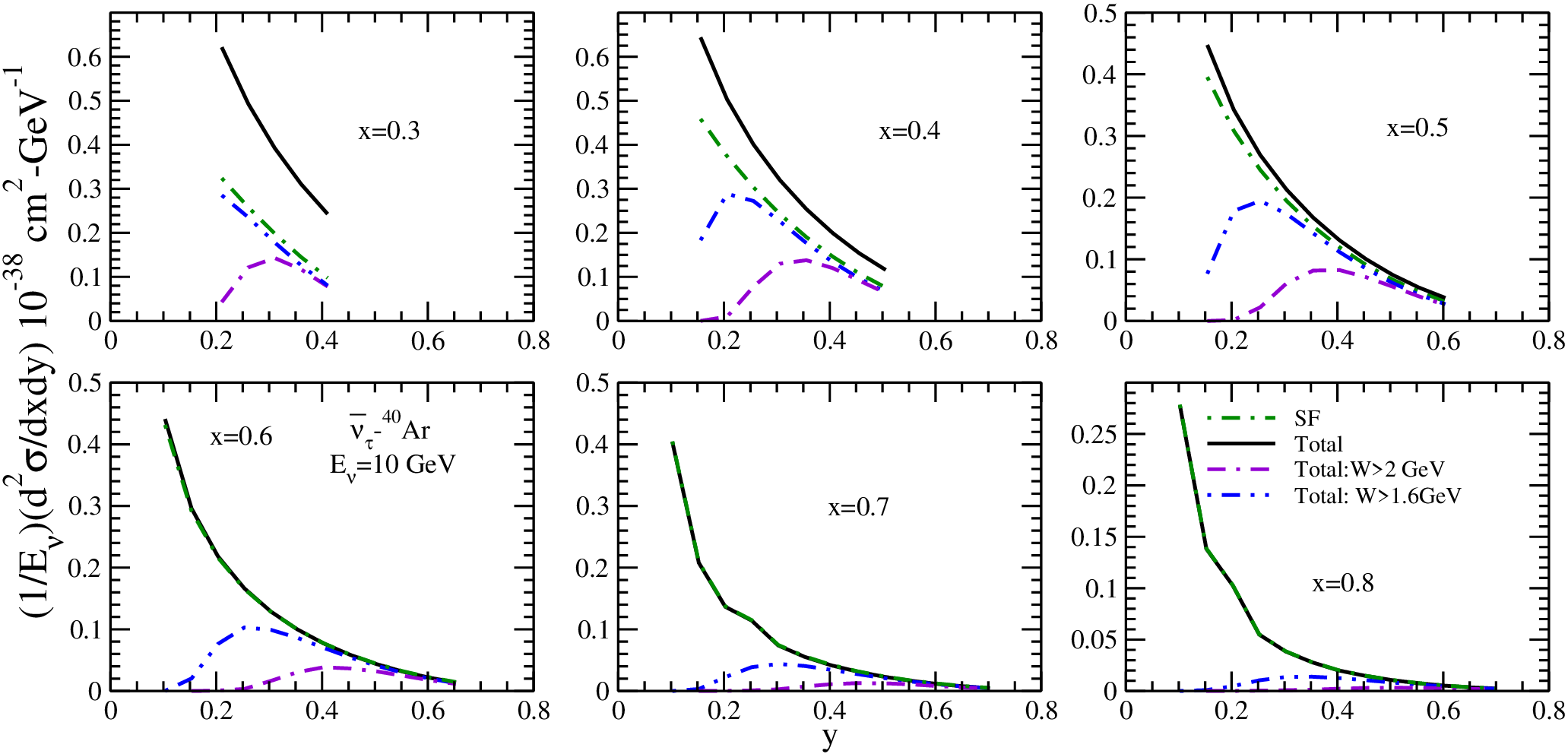}
\caption{Results for the double differential scattering cross section, $\frac{1}{E_\nu}\frac{d^2\sigma_A}{dx dy}$ vs $y$, are shown at the different values of
$x$ for $E_\nu=10$ GeV in $\bar\nu_\tau-^{40}Ar$. The lines and symbols have the same meaning as in Fig.~\ref{fig_d2sig5gev}.}
\label{fig_d2sig10geva}
 \end{figure}

\begin{figure}
 \includegraphics[height=10 cm, width=.95\textwidth]{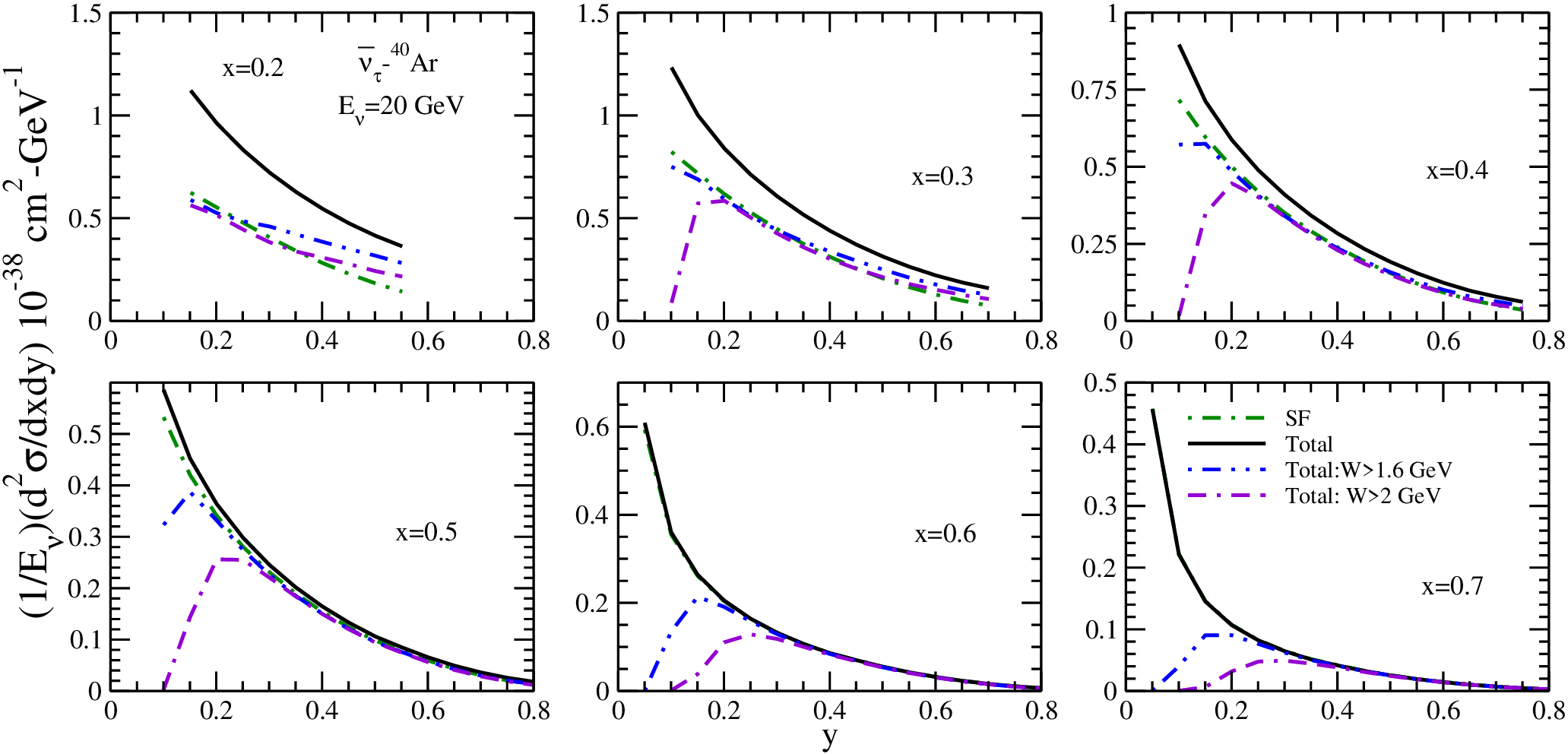}
\caption{Results for the double differential scattering cross section, $\frac{1}{E_\nu}\frac{d^2\sigma_A}{dx dy}$ vs $y$, are shown at the different values of
$x$ for $E_\nu=20$ GeV in $\bar\nu_\tau-^{40}Ar$. The lines and symbols have the same meaning as in Fig.~\ref{fig_d2sig5gev}.}
\label{fig_d2sig20geva}
 \end{figure}

In Figs.~\ref{fig_d2sig5geva}, \ref{fig_d2sig10geva} and \ref{fig_d2sig20geva}, the results for the antineutrino ($\bar\nu_\tau$) induced reaction on the argon nuclear 
target have been presented. These results are shown for $\frac{1}{E_\nu}\frac{d^2\sigma_A}{dx dy}$ vs $y$. The qualitative behavior of the differential scattering cross section and its 
modifications due to the nuclear medium effects is similar to that  observed in the case of $\nu_\tau-^{40}Ar$ induced DIS process (Figs.~\ref{fig_d2sig5gev}, \ref{fig_d2sig10gev}
and \ref{fig_d2sig20gev}). However, quantitatively the nuclear medium effects on the $\bar\nu_\tau-^{40}Ar$ cross sections are found to be larger at low $x$ as
compared to the case of $\nu_\tau-^{40}Ar$ cross sections.
For example, on 
comparing the results obtained with the full model (''Total") and the results obtained only using the spectral function(``SF"), we find that the $\nu_\tau-^{40}Ar$ cross section gets 
enhanced by $30\%(1\%)$ at $E_\nu=10$ GeV, $y=0.3$ and $x=0.3(0.6)$, while $\bar \nu_\tau-^{40}Ar$ cross section gets enhanced by  $50\%(<1\%)$.
To study the effect of the CM energy cut on $\bar\nu_\tau-^{40}Ar$ scattering cross sections in Figs.\ref{fig_d2sig5geva}, \ref{fig_d2sig10geva} and \ref{fig_d2sig20geva} we have 
compared the results when we apply no cut on the CM energy (solid line) and when a cut of 2GeV ($W>2.0GeV$) is applied (double dash-dotted line). We find
a suppression of about $41\%(86\%)$ in the $\nu_\tau-^{40}Ar$ and $63\%(87\%)$ in the $\bar \nu_\tau-^{40}Ar$ scattering cross sections at $E_\nu=10GeV$, $y=0.3$ and $x=0.3(0.6)$.

   \begin{figure}[h]
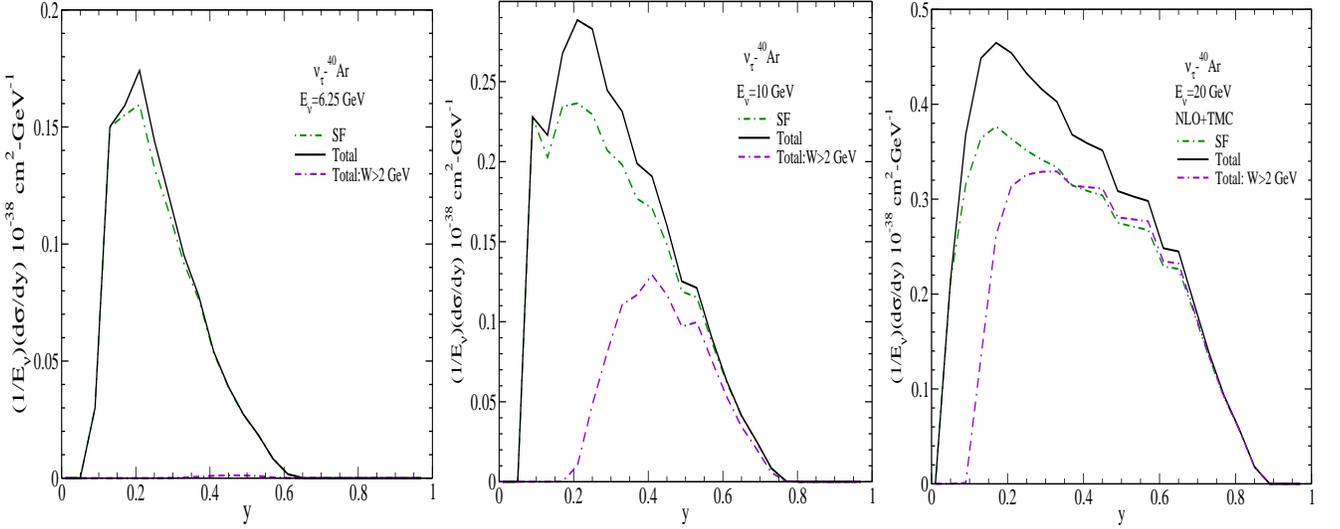

     \includegraphics[height=7 cm, width=0.32\textwidth]{dsigy_argon_nutau_625gev_v1.eps}
 \includegraphics[height=7 cm, width=0.32\textwidth]{dsigy_argon_nutau_10gev_v1.eps}
  \includegraphics[height=7 cm, width=0.32\textwidth]{dsigy_argon_nutau_20gev_v1.eps}
\caption{$\frac{1}{E_\nu}\left(\frac{d\sigma_A}{dy}\right)$ vs $y$ is shown for $E_\nu=6.25$ GeV (left panel), 10 GeV (middle panel) and 20 GeV (right panel) in $\nu_\tau-^{40}Ar$. 
`SF' corresponds to the results only with the spectral 
function, and `Total' corresponds to the results of full model, where the additional contribution from the mesonic clouds and the shadowing
effects are taken into account. Results are obtained in the four flavor MSbar-scheme (massless quarks viz. $u$, $d$, $s$ and massive $c$ quark) by incorporating the 
TMC~\cite{Kretzer:2003iu} and HT~\cite{Dasgupta:1996hh} effects at NLO. For the numerical calculations MMHT nucleon PDFs parameterization~\cite{Harland-Lang:2014zoa} has been used.
Dash-dotted and solid lines represent the results of cross section only with the 
spectral function and with the full model, respectively without having any cut on the CM energy $W$. Double dash-dotted line shows the results for the 
full model with a cut of $W>2.0$ GeV on the CM energy.}
\label{fig_dsignu}
 \end{figure}
    \begin{figure}[h]
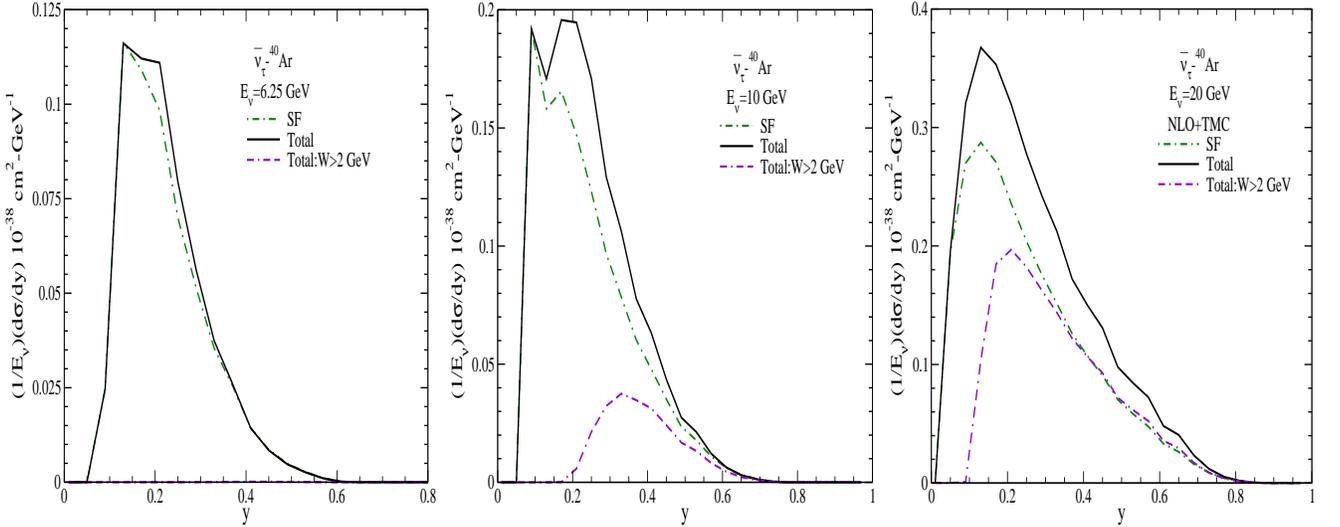

    \includegraphics[height=7 cm, width=0.32\textwidth]{dsigy_argon_nutaubar_625gev_v1.eps}
 \includegraphics[height=7 cm, width=0.32\textwidth]{dsigy_argon_nutaubar_10gev_v1.eps}
  \includegraphics[height=7 cm, width=0.32\textwidth]{dsigy_argon_nutaubar_20gev_v1.eps}
\caption{$\frac{1}{E_\nu}\left(\frac{d\sigma_A}{dy}\right)$ vs $y$ for $E_\nu=6.25$ GeV (left panel), 10 GeV (middle panel) and 20 GeV (right panel) in $\bar\nu_\tau-^{40}Ar$. 
The lines and symbols have the same meaning as in Fig.~\ref{fig_dsignu}.}
\label{fig_dsiganu}
 \end{figure}
 
 In Figs.~\ref{fig_dsignu} and \ref{fig_dsiganu}, we have presented the results for $\frac{1}{E_\nu}\left(\frac{d\sigma_A}{dy}\right)$ vs $y$, respectively, 
 for $\nu_\tau$ and $\bar\nu_\tau$ induced DIS 
 processes by integrating  $\frac{1}{E_\nu}\frac{d^2\sigma_A}{dxdy}$ over $x$ in the kinematic region as defined in Eq.~\ref{xydef}. These results are obtained
 for $6.25 ~\textrm{GeV} \le E_\nu \le 20~\textrm{GeV}$ at NLO with the TMC and HT effects in the four flavor MSbar scheme. It is important to notice that scattering cross section peaks in the region of low $y(\sim 0.2)$ 
 irrespective of the incoming neutrino energy. 
 When the results obtained using only the spectral function ``SF" (dash dotted line) are compared with the results obtained using the full model ``Total" (solid line), we find an enhancement
 of $15\%(5\%)$ in the $\nu_\tau-^{40}Ar$ and $25\%(14\%)$ in the $\bar \nu_\tau-^{40}Ar$  cross section at $E_\nu=10$ GeV and $y=0.3(0.5)$.
 The effect of charm mass has also been studied and found it to be negligible in the overall energy region of present interest (not shown here explicitly).
  The results in these figures are also compared when there is no cut (solid line) on the center of mass energy and when a cut of $W>2$ GeV is applied (double dashed-dotted line),
 considering the region of $Q^2\ge 1$ GeV$^2$ and $W\ge 2$ GeV to be the region of safe DIS~\cite{SajjadAthar:2020nvy, Athar:2020kqn}. From the figure, it
 may be noticed that the results of the differential cross section obtained with a cut of $W\ge 2$ GeV is very small at $E_\nu=6.25$ GeV, i.e. in the safe DIS region 
 at this energy value, the $\tau-$lepton production is small. While at higher energies viz. $E_\nu=10$ GeV and $E_\nu=20$ GeV, there is significant contribution of tau lepton events 
 which results an enhancement in the differential cross section. From a quantitative analysis we find that due to the effect of $W>2$ GeV cut the results of differential cross sections at $E_\nu=10$GeV get
 reduced by $67\%(22\%)$ for the neutrino induced process and by $75\%(38\%)$ for the antineutrino process at $y=0.3(0.5)$.
  The differential scattering cross section for $\bar\nu_\tau-^{40}Ar$ interaction is found to be $36\% (82\%)$ smaller from 
  the one obtained for $\nu_\tau-^{40}Ar$ scattering at $y=0.2 (0.5)$ for $E_\nu=6.25$ GeV. This reduction is found to be energy 
  dependent and becomes $32\% (78\%)$ at $E_\nu=10$ GeV and $30\%(68\%)$ at $E_\nu=20$ GeV for $y=0.2 (0.5)$. 

\subsection{Summary and conclusions}
In this work, we have presented the results for the nuclear structure functions($F_{iA}(x,Q^2)$, $i=1-5$) 
and the double ($\frac{d^2\sigma}{dxdy}$) and single ($\frac{d\sigma}{dy}$) differential scattering cross sections
for the charged current $\nu_\tau / {\bar\nu}_\tau-^{40}$Ar deep inelastic scattering by incorporating perturbative and nonperturbative effects including the nuclear medium effect.

This is the first work which has explicitly dealt with the nuclear medium effects in the evaluation of $F_{4A}(x,Q^2)$ and $F_{5A}(x,Q^2)$ structure functions. These structure 
functions become significant for the tau leptons produced in the charged current $\nu_\tau / {\bar\nu}_\tau$ interactions from the nuclear target. In the
evaluation of nuclear structure functions $F_{iA}(x,Q^2);~(i=1-5)$, nucleon structure functions $F_{iN}(x,Q^2);~(i=1-5)$, are taken as input and then convoluted
with the spectral function of the nucleons in the nuclear medium, to take into account the Fermi motion, binding energy and nucleon correlation effects. At the nucleon level, we have assumed 
Callan-Gross relation ($F_{2}(x)=2 x F_{1}(x)$) and Albright-Jarlskog ($F_{2}(x)=2 x F_{5}(x)$) relation. However, in the case of nuclei all the nuclear structure functions $F_{iA}(x,Q^2);~(i=1-5)$
 were evaluated independently. In addition to that, we have considered the mesonic contributions and shadowing effects while evaluating the nuclear 
structure function. Both of these effects are included in $F_{1A}(x,Q^2)$ and $F_{5A}(x,Q^2)$ by using the Callan-Gross and Albright-Jarlskog relations. Furthermore, in $F_{3A}(x,Q^2)$ there is no mesonic effect and 
only shadowing effect contributes along with the nucleon spectral function while in $F_{4A}(x,Q^2)$ both the shadowing effect and mesonic contributions are absent. The kinematic region
in which these studies have been done are not only important to 
the DUNE experiment but also to the HyperK and  IceCube experiments as well as to the atmospheric neutrino experiments~\cite{Aoki:2019jry, SHiP:2018xqw, DiCrescenzo:2016irr, Agafonova:2018auq, Abe:2012jj, Li:2017dbe}.

Our findings are as follows:

 \begin{itemize}
  \item The inclusion of perturbative and nonperturbative effects is quite important in the evaluation of the nucleon structure functions as 
  well as in the evaluation of the differential scattering cross sections. 
  \item The nuclear structure functions obtained only with the spectral function is suppressed from the free nucleon case in the entire region of $x$. 
  However, with the increase in $Q^2$, it has been observed that the suppression in the nuclear structure functions due to nuclear medium effects becomes small.
  \item  When the mesonic contributions are included, we find an enhancement in the nuclear structure functions $F_{1A}(x,Q^2)$, $F_{2A}(x,Q^2)$ and $F_{5A}(x,Q^2)$ in the low 
  and intermediate region of $x$. 
  We observe that the mesonic contribution is dominant in the region of $0.2\le x \le0.6$, and decreases with the increase in $Q^2$. 
  \item We find that at low energy the double differential scattering cross section $\frac{d^2\sigma_A}{dxdy}$ contributes only in the intermediate and high region of $x$ for low and mid
  range of $y$. The results obtained with the center of mass energy cut are found to be very small at $E_\nu=6.25$ GeV. Although at higher neutrino energies ($E_\nu=10$ GeV and 20 GeV) 
  scattering cross section get enhanced but even at these energies we observe that in the region of low $y$ there is significant suppression in the results due to the effect of CM energy cut.
It implies that the definition of sharp kinematic limits for the safe DIS region is quite important in order to avoid the contribution coming from the inelastic region to be calculated 
using DIS formalism.
  \item For antineutrino induced process the scattering cross section gets reduced as compared to the case of neutrino induced process which was expected. However, the qualitative behavior
  of lepton mass effect, center of mass energy cut, massive charm quark and nuclear medium effects are found to be similar.
  \item  The effect of $\tau-$lepton mass is found to be significant at low energies in the region of low and intermediate $x$. However, with the increase in energy the lepton mass effect
  gradually decreases.
  \item The differential scattering cross section $\frac{d\sigma}{dy}$ peaks in the low $y$ region irrespective of the (anti)neutrino energies. 
 \end{itemize}

 Thus to conclude these theoretical results describing the nuclear medium effects in various regions of Bjorken $x$ and inelasticity $y$ for $\nu_\tau(\bar\nu_\tau)-^{40}Ar$ scattering, 
 would be helpful to understand the experimental results from DUNE. Furthermore, these results are also important in understanding the results from the experiments being performed using 
 atmospheric neutrinos.
\section*{Acknowledgment}   
F. Zaidi is thankful to the Council of Scientific \& Industrial Research (CSIR), India, for providing the research associate fellowship with 
award letter no. 09/112(0622)2K19 EMR-I.
M. S. A. is thankful to the Department of Science and Technology (DST), Government of India for providing 
financial assistance under Grant No. SR/MF/PS-01/2016-AMU/G. I.R.S. acknowledges support from project PID2020-114767GB-I00 funded by MCIN/AEI/10.13039/501100011033, from 
project A-FQM-390-UGR20 funded by FEDER/Junta de Andalucia-Consejeria de Transformacion Economica, Industria, Conocimiento y Universidades, and by Junta de Andalucia (grant No FQM-225).

\end{document}